\documentclass[twocolumn]{aastex63}

 \usepackage{todonotes}
 \usepackage{tabularx}
 \usepackage{soul}
 \usepackage{multirow}
 \DeclareUnicodeCharacter{2212}{-}

\newcommand{\HII}{\ion{H}{2}}

\submitjournal{ApJ}

\shortauthors{Santos et al.}
\graphicspath{{./}{figures/}}

\begin{document}

\title{A spectral survey of CH$_3$CCH in the Hot Molecular Core G331.512-0.103}

\correspondingauthor{Julia C. Santos}
\email{santos@strw.leidenuniv.nl}

\author[0000-0002-3401-5660]{Julia C. Santos}
\affiliation{Instituto de Astronomia, Geofísica e Ciências Atmosféricas; Universidade de São Paulo, Rua do Matão, 1226, São Paulo, SP, Brazil}
\affiliation{Laboratory for Astrophysics, Leiden Observatory, Leiden University, PO Box 9513, 2300 RA Leiden, The Netherlands}

\author{Leonardo Bronfman}
\affiliation{Departamento de Astronomía; Universidad de Chile, Casilla 36-D, Santiago de Chile, Chile}

\author[0000-0001-9381-7826]{Edgar Mendoza}
\affiliation{Instituto de Astronomia, Geofísica e Ciências Atmosféricas; Universidade de São Paulo, Rua do Matão, 1226, São Paulo, SP, Brazil}
\affiliation{Observatório do Valongo; Universidade Federal do Rio de Janeiro, Ladeira do Pedro Antônio, 43, Rio de Janeiro, RJ, Brazil}

\author[0000-0002-8969-0313]{Jacques R. D. Lépine}
\affiliation{Instituto de Astronomia, Geofísica e Ciências Atmosféricas; Universidade de São Paulo, Rua do Matão, 1226, São Paulo, SP, Brazil}

\author{Nicolas U. Duronea}
\affiliation{Instituto de Astrofísica de La Plata (UNLP-CONICET), La Plata, Argentina}
\affiliation{Faculdad de Ciencias Astronómicas y Geofísicas, Universidad Nacional de La Plata, Paseo del Bosque s/n, 1900, La Plata, Argentina}

\author[0000-0003-0709-708X]{Manuel Merello}
\affiliation{Departamento de Astronomía; Universidad de Chile, Casilla 36-D, Santiago de Chile, Chile}

\author{Ricardo Finger}
\affiliation{Departamento de Astronomía; Universidad de Chile, Casilla 36-D, Santiago de Chile, Chile}




\begin{abstract}

A spectral survey of methyl acetylene (CH$_3$CCH) was conducted toward the hot molecular core/outflow  G331.512-0.103. Our APEX observations allowed the detection of 41 uncontaminated rotational lines of CH$_3$CCH in the frequency range between 172--356 GHz. Through an analysis under the local thermodynamic equilibrium assumption, by means of rotational diagrams, we determined $T_{exc}$=50$\pm$1 K, $N$(CH$_3$CCH)$=$(7.5$\pm$0.4)$\times$10$^{15}$ cm$^{−2}$, $X$[CH$_3$CCH/H$_2$]$\approx$(0.8--2.8)$\times10^{-8}$ and $X$[CH$_3$CCH/CH$_3$OH]$\approx$0.42$\pm$0.05 for an extended emitting region ($\sim$10\arcsec). The relative intensities of the $K$=2 and $K$=3 lines within a given $K$-ladder are strongly negatively correlated to the transitions’ upper J quantum-number ($r$=-0.84). Pure rotational spectra of CH$_3$CCH were simulated at different temperatures, in order to interpret this observation. The results indicate that the emission is characterized by a non-negligible temperature gradient with upper and lower limits of $\sim$45 and $\sim$60 K, respectively. Moreover, the line widths and peak velocities show an overall strong correlation with their rest frequencies, suggesting that the warmer gas is also associated with stronger turbulence effects. The $K$=0 transitions present a slightly different kinematic signature than the remaining lines, indicating that they might be tracing a different gas component. We speculate that this component is characterized by lower temperatures, and therefore larger sizes. Moreover, we predict and discuss the temporal evolution of the CH$_3$CCH abundance using a two-stage zero-dimensional model of the source constructed with the three-phase Nautilus gas-grain code.

\end{abstract}



\section{Introduction} \label{sec:intro}

Star-forming regions play a key role in building the complex inventory of chemical species detected in astronomical environments, which in turn serve as powerful tools to study their surroundings. Through the observation of molecular emission lines, it is possible to constrain both the physical conditions and the chemical evolution of circumstellar sources, shedding light on the formation processes that connect different evolutionary stages of star-formation. Methyl acetylene (CH$_3$CCH), also known as propyne, is a particularly good temperature probe, and has therefore been widely explored toward various star-forming objects (e.g., \citealt{Fontani2002,Bisschop2007,Fayolle2015,Giannetti2017,Andron2018,Calcutt2019}).

Massive stars ($L > 10^3 L_\odot$; $M > 8M_\odot$) greatly affect their surroundings through their feedback and the production of heavy elements. The formation process of those stars, however, is still less well-understood in comparison to the low-mass counterparts. Observational disadvantages, such as their complex cluster environments ($n_{\star} \gtrsim 100$ pc$^{-3}$) and large distances involved ($d\geq 1$ kpc), together with a considerably shorter evolutionary timescale ($t_{KH}\leq 10^4$ yr for O-type stars), substantially impair the development of a solid massive star-formation paradigm \citep{Garay1999,Zinnecker2007,Tan2014,Krumholz2015,Silva2017,Rosen2020}.

Despite the lack of a thoroughly defined evolutionary sequence, some precursors of high-mass stars are well established. One example are Hot Molecular Cores (HMCs), which are one of the first manifestations of massive star formation \citep{Cesaroni2005}. They are characterized by relatively high temperatures ($>$100 K) and high densities ($n_{H_2}$$\sim$$10^5$--$10^8$ cm$^{-3}$), linked with a compact ($<$0.1 pc), luminous ($>$10$^4~L_{\odot}$) and massive ($\sim$10--1000 M$_{\odot}$) molecular core \citep{Heaton1989, Gomez1995, Cesaroni1998, Cesaroni2005, Bonfand2019}. HMCs are associated with a rich molecular emission spectrum, which carries information on their chemical and physical properties, as well as their morphology and evolutionary stage (e.g., \citealt{Caselli1993,Comito2005,Herbst2009,Allen2018,Bonfand2019,Jorgensen2020,Gieser2021}).

\subsection{The source: G331.512-0.103}

G331.512-0.103 (henceforth G331) is a massive and energetic hot molecular core/outflow system. It is located within the millimeter source G331.5-0.1 (see \citealt{Merello2013} and references therein), at the tangent point of the Norma spiral arm ($\sim 7.5$ kpc) \citep{Bronfman1985,Bronfman2008}, and which consists of a singularly extended and luminous complex at the center of a Giant Molecular Cloud (GMC). The parent GMC is located at the peak region of the southern molecular and massive-star-formation rings (i.e., the peaks of the azimuthally averaged radial distributions of molecular gas and of regions of massive star formation, \citealt{Bronfman2000}). Likewise, it is one of the most massive and active star forming clouds in the southern Galaxy \citep{Garcia2014}. Observational evidence indicates that a Young Massive Stellar Object (YMSO) at the center of G331 drives a powerful outflow with $\sim 55 ~M_\odot$ of mass and a momentum of $\sim 2.4 \times 10^3 ~M_\odot$ km s$^{-1}$ \citep{Bronfman2008}, with lobes closely aligned with the line of sight. Observations with the Atacama Large Millimeter/submillimeter Array (ALMA) of the SiO(8-7), H$^{13}$CO$^+$(4-3), HCO$^+$(4-3) and CO(3-2) transitions revealed an expanding bubble geometry driven by stellar winds, which probably arises from the protostar confined in a compact \HII ~region of $\sim$5 arcsec (projected size of $\sim$0.2 pc) \citep{Merello2013b}. Its emission spectrum exhibits great chemical lavishness, from common radicals and carbon-chain molecules to prebiotic and complex organic species
\citep{Merello2013b, Merello2013,Mendoza2018,Duronea2019,Hervias-Caimapo2019, Canelo2021}.
All those factors make G331 an exquisite source for astrochemical investigations by means of spectral surveys in sub-mm wavelenghts.

\subsection{The molecule: Methyl acetylene}

More than 200 molecules have been hitherto detected toward the interstellar and circumstellar media (e.g., \citealt{McGuire2018a}). Methyl acetylene (CH$_3$CCH) has been detected in a wide variety of sources. It has been proven to be a reliable tracer of physical conditions, such as temperature and density, and has been extensively studied toward star-forming regions (e.g., \citealt{Buhl1973, Lovas1976, Hollis1981,Kuiper1984,Taniguchi18,Guzman2018,Bogelund2019,Calcutt2019}). It has also been observed toward extragalactic sources, such as M 82, NGC 253, and NGC 1068 \citep{mauersberger1991,Qiu2020}, and toward a planetary nebula \citep{Schmidt2019}. 

The rotational transitions of CH$_3$CCH are characterized by two quantum numbers, namely the total angular momentum ($J$) and its projection on the principal symmetry axis ($K$) \citep{Townes1975,Muller2000}. Because it is a symmetric rotor, CH$_3$CCH presents many transitions that are closely spaced in frequency: the so-called K-ladders. The capability to observe many lines in the same bandwidth reduces calibration uncertainties and yields more precise predictions. Moreover, due to its small electric dipole moment ($\mu=0.75$ D) \citep{Dubrule1978}, line thermalization occurs at densities as low as $\sim$10$^4$ cm$^{-3}$ (e.g., \citealt{Bergin1994,Fontani2002,Molinari2016}). Transitions with $\Delta K \neq 0$ are forbidden, and therefore the relative population of different K-ladders are dictated by collisions. As a result, CH$_3$CCH acts as an excellent temperature probe. Indeed, this molecule's temperature sensitivity makes it a reliable tracer of physical conditions and passive heating \citep{Giannetti2017, Molinari2016}.

Aside from observational advantages, the intrinsic spectroscopic properties of CH$_3$CCH can directly affect the observed spectra and therefore can give powerful insights on the physics of the emitting gas. Rotational spectroscopy is a conspicuously fruitful technique to infer information on the molecular species and its environment (e.g., \citealt{Domenicano1992,Winnewisser2003,Grubbs2010}), and is nowadays highly assisted by quantum-chemistry (e.g., \citealt{Puzzarini2012a,Puzzarini2012b,Cernicharo2015,McGuire2016,Cazzoli2016,Cerqueira2020,Santos2020}). The rotational Hamiltonian is described in terms of the zeroth-order rotational constants (A, B, and C), and the higher-order centrifugal-distortion constants (e.g., D$_J$, D$_{JK}$, D$_{K}$, H$_{J}$, H$_{JK}$, H$_{KJ}$ and H$_{K}$, for symmetric tops). These constants can be measured from laboratory experiments guided by theoretical simulations, which in turn play a fundamental role in line assignments of radioastronomical observations (e.g., \citealt{Belloche2014,Coutens2016,Melli2018,Belloche2019}).

In this work, we conducted a spectral survey of CH$_3$CCH toward G331, resulting in the detection of 41 uncontaminated lines. The goal of this work is to analyze the excitation conditions of methyl acetylene in this source, which gives information on the chemical and physical conditions of the gas and consequently helps unveiling the early stages of massive star-formation. Quantum-chemical properties associated with this molecule are especially relevant to explore the small-scale structure of the source. In section \ref{sec:obs}, we delineate the observational procedure. In section \ref{sec:results}, we present our results on the line identifications and radiative analysis. Those results are then discussed in section \ref{sec:discussion}. The observed abundance is discussed in the context of a chemical model of our source in section \ref{sec:modelling}. Finally, our conclusions and perspectives are presented in section \ref{sec:conclusions}.


\section{Observations} \label{sec:obs}

The observations were obtained  with the Atacama Pathfinder EXperiment (APEX) telescope \citep{gu06} using the single-point mode toward the coordinates of the source \citep{Bronfman2008} RA, DEC\ (J2000) = 16$^h$12$^m$10.1$^s$, $-$51$^{\circ}$28$^{\prime}$38.1$^{\prime\prime}$. The Swedish-ESO PI Instrument (SEPIA180; \citealt{Belitsky2018}) was used, together with APEX-1 and APEX-2 receivers of the  Swedish Heterodyne Facility Instrument (SHeFI; \citealt{vas08}), to observe nine frequency setups within the intervals 170--205~GHz and 222--307~GHz, respectively. As backend for the APEX-1 and APEX-2 receivers, the eXtended bandwidth Fast Fourier Transform Spectrometer2 (XFFTS2) was used, which consists of two units with 2.5 GHz bandwidth divided into 32768 channels each. SEPIA180 is a dual-polarization sideband-separated (2SB) receiver, which is able to observe two 4 GHz band, separated by 12 GHz, simultaneously. Thus, it covers 100\% of the SEPIA ALMA Band 5 receiver channel IF band (159--211 GHz). The spectral resolution, corresponding to velocity resolutions, was obtained between 0.06 and 0.13~km~s$^{-1}$ for a noise level of $\sim$30--50 mK. While all analyses were performed with the original data, for further clarity and uniformity in their graphical depiction, the resolution of the spectra exhibited in the present work were degraded to a common value of 1~km~s$^{-1}$. The original intensity, obtained in a scale of antenna temperature corrected for atmospheric attenuation ($T_A$), was converted to the main-beam temperature ($T_{mb}$) scale using the main-beam efficiencies $\eta_{mb}$~= 0.80 for SEPIA180\footnote{\url{http://www.apex-telescope.org/telescope/efficiency/}}, and $\eta_{mb}$~= 0.75 and 0.73 for APEX-1 and APEX-2, respectively\footnote{\url{https://www.apex-telescope.org/telescope/efficiency/index.php.old}} (e.g. \citealt{Canelo2021}). The Half Power Beam Width (HPBW) values vary between $\sim$17--39\arcsec.\footnote{\url{https://www.apex-telescope.org/instruments/}} We adopted a calibration uncertainty of about 10\% \citep{dum2010}.

The data reduction was carried out using the CLASS package of the GILDAS software\footnote{\url{https://www.iram.fr/IRAMFR/GILDAS/}} \citep{Pety2005,Gildas2013}. First-degree polynomial baselines were removed from each individual scan, which were subsequently averaged into a final spectrum. Line identifications were performed using the Weeds extension of CLASS, in combination with spectroscopic databases such as NIST\footnote{\url{https://physics.nist.gov/cgi-bin/micro/table5/start.pl}} Recommended Rest Frequencies for Observed Interstellar Molecular Microwave Transitions \citep{Lovas2004}, CDMS\footnote{\url{https://cdms.astro.uni-koeln.de/}} \citep{Muller2001,Muller2005,end2016}, JPL\footnote{\url{https://spec.jpl.nasa.gov/}} \citep{pic1998}, and Splatalogue\footnote{\url{https://www.cv.nrao.edu/php/splat/}}. The establishment of a molecular identification warrants the satisfaction of a set of standard requirements \citep{Snyder2005}. Accordingly, the following criteria were used in order to confirm a detection: the peak frequencies of the observed lines should be consistent with the systemic velocity of G331 ($V_{lsr}\sim-90$ km s$^{-1}$), the intensities of the observed lines should surpass the threshold of 3 rms noise, and lines predicted through Local Thermodynamic Equilibrium (LTE) modeling should agree with the observations. The radiative studies, including the optical depth estimation, were based on analyses carried out with the CASSIS software\footnote{\url{http://cassis.irap.omp.eu/}} \citep{Vastel2015}---assuming LTE. CASSIS has been developed by IRAP-UPS/CNRS.


\section{Results} \label{sec:results}

\subsection{Line analysis of CH$_3$CCH}\label{sec:3.1}

\begin{figure*}[htb]
\centering
\includegraphics[scale=0.35]{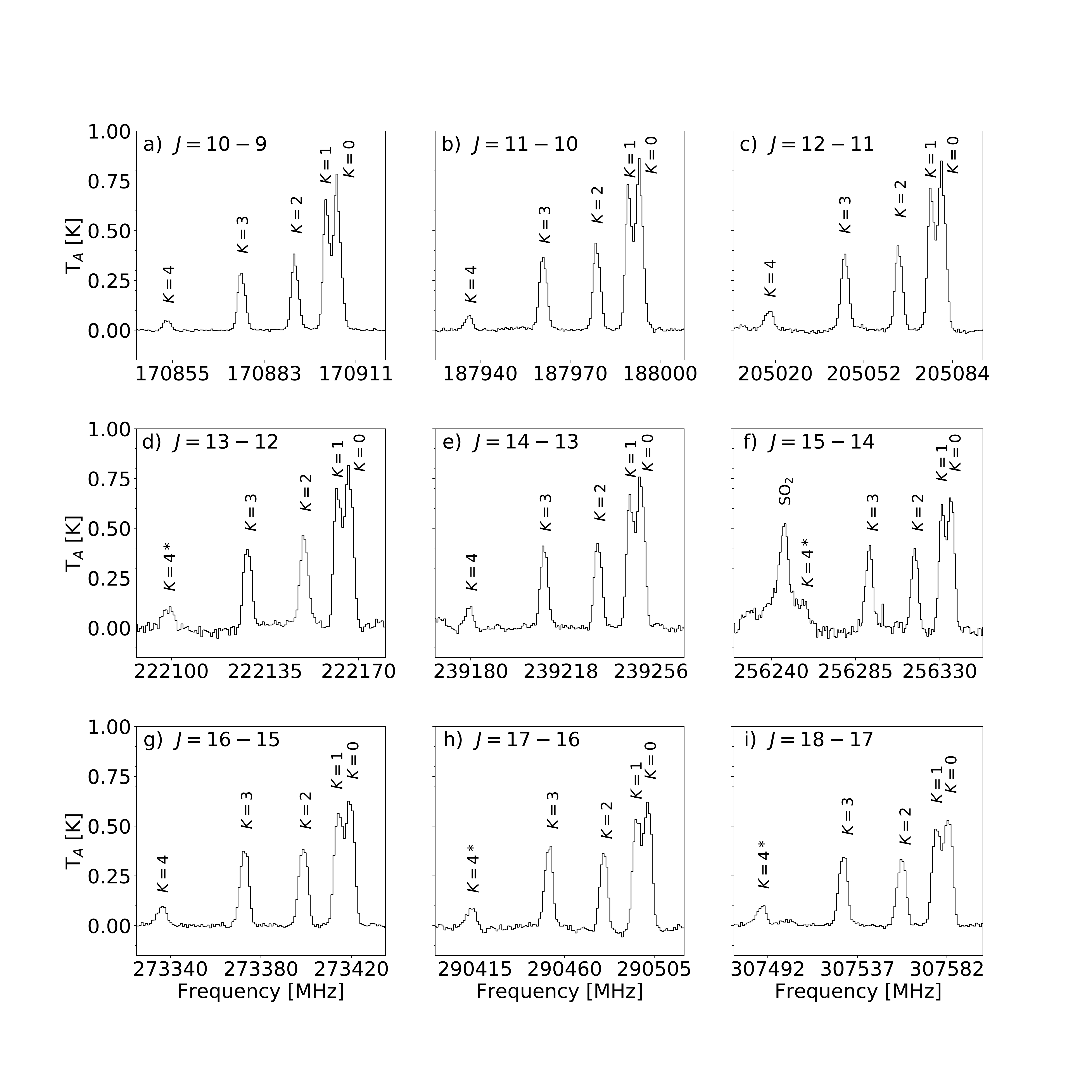}
\caption{Spectral lines of CH$_3$CCH identified toward G331 across the frequency range of 170.850--307.590~GHz. Each panel displays the $K$-ladder lines for the rotational transitions a) $J=10_K-9_K$, b) $J=11_K-10_K$, c) $J=12_K-11_K$, d) $J=13_K-12_K$, e) $J=14_K-13_K$, f) $J=15_K-14_K$, g) $J=16_K-15_K$, h) $J=17_K-16_K$ and i) $J=18_K-17_K$. The asterisk (${\ast}$) indicates lines below the 3$\sigma$ threshold. Linear baselines were subtracted from each spectrum, and the spectral resolution was resampled to a common channel of approximately 1~km~s$^{-1}$.}
\label{fig:ch3cch_lines}
\end{figure*}

We detected a total of 41 lines of CH$_3$CCH spread across the spectral band, from $\sim$170.84 GHz to $\sim$307.60 GHz. Considering that CH$_3$CCH exhibits a K-ladder spectral signature, in Figure \ref{fig:ch3cch_lines} we have displayed all the observed K-ladder structures from the $J$=10--9 to $J$=18--17 rotational levels. 

Spectroscopic parameters obtained through Gaussian fittings to the lines are summarized in Table \ref{table:CH3CCH}. The fittings were performed to the spectra at their full resolution of $\sim$0.06--0.13 km s$^{-1}$.

\begin{table*}[htb!]
\caption{Spectral lines of CH$_3$CCH observed toward G331 and their parameters obtained from Gaussian fits. $^{\ast}$Lines below the 3$\sigma$ threshold. }
\label{table:CH3CCH}      
\centering
\begin{tabularx}{0.92\linewidth}{c c c c c l c c c c c}
\hline\hline                        
\multicolumn{5}{c}{Quantum Numbers} & Frequency & $E_u$ & $A_u$ & $\int$T$_A$~dv &  $V_{lsr}$ & FWHM\\
$J_u$ & $K_u$ & -- & $J_l$  & $K_l$ & (MHz) & (K) & (10$^{-5}$ s$^{-1}$) & (K km s$^{-1}$) & (km s$^{-1}$) & (km s$^{-1}$) \\
\hline  
10	&   4	&&  9	&   4	&   170853.576	        &   160.7	&   1.43	&   0.24	$\pm$   0.01    &   -90.02	$\pm$   0.09	&   4.4	    $\pm$   0.2\\
10	&   3   &&	9	&   3	&   170876.410	        &   110.1	&   1.55	&   1.43    $\pm$   0.01	&   -89.89  $\pm$	0.02	&   4.57    $\pm$	0.06\\
10	&   2	&&  9	&   2	&   170892.726	        &   74.0	&   1.63	&   1.75	$\pm$   0.02	&   -89.77	$\pm$   0.03	&   4.41	$\pm$   0.06\\
10	&   1	&&  9	&   1	&   170902.518	        &   52.3	&   1.68	&   3.04	$\pm$   0.06	&   -89.74	$\pm$   0.04	&   4.4	    $\pm$   0.1\\
10	&   0	&&  9	&   0	&   170905.783	        &   45.1	&   1.70	&   3.95	$\pm$   0.04	&   -89.34	$\pm$   0.02	&   4.99	$\pm$   0.06\\
\hline
11	&   4	&&  10	&   4	&   187936.230	        &   169.7	&   1.97	&   0.38	$\pm$   0.01	&   -90.40	$\pm$   0.08	&   4.76	$\pm$   0.2\\
11	&   3	&&  10	&   3	&   187961.342	        &   119.2	&   2.10	&   1.87	$\pm$   0.02	&   -89.96	$\pm$   0.02	&   4.68	$\pm$   0.05\\
11	&   2	&&  10	&   2	&   187979.286	        &   83.0	&   2.20	&   2.07	$\pm$   0.02	&   -89.84	$\pm$   0.02	&   4.45	$\pm$   0.06\\
11	&   1	&&  10	&   1	&   187990.055	        &   61.4	&   2.25	&   3.53	$\pm$   0.07	&   -89.83	$\pm$   0.04	&   4.6	    $\pm$   0.1\\
11	&   0	&&  10	&   0	&   187993.645	        &   54.1	&   2.27	&   4.43	$\pm$   0.04	&   -89.39  $\pm$   0.02	&   5.06	$\pm$   0.06\\
\hline
12	&   4	&&  11	&   4	&   205018.114	        &   179.6	&   2.63	&   0.48	$\pm$   0.01	&   -89.89	$\pm$   0.07	&   5.0	    $\pm$   0.2\\
12	&   3	&&  11	&   3	&   205045.501	        &   129.0	&   2.77	&   1.97	$\pm$   0.02	&   -89.99	$\pm$   0.02	&   4.80	$\pm$   0.06\\
12	&   2	&&  11	&   2	&   205065.070          &	92.9	&   2.88	&   2.04	$\pm$   0.02	&   -89.86	$\pm$   0.02	&   4.52	$\pm$   0.05\\
12	&   1	&&  11	&   1	&   205076.816	        &   71.2	&   2.94	&   3.51	$\pm$   0.06	&   -89.94	$\pm$   0.04	&   4.6	    $\pm$   0.1\\
12	&   0	&&  11	&   0	&   205080.732	        &   64.0	&   2.96	&   4.46	$\pm$   0.05	&   -89.43	$\pm$   0.03	&   5.18	$\pm$   0.07\\
\hline
13	&   4	&&  12	&   4	&   222099.153$^{\ast}$	&   190.2	&   3.42	&   0.72	$\pm$   0.04	&   -89.9	$\pm$   0.2	    &   6.6	    $\pm$   0.4\\
13	&   3	&&  12	&   3	&   222128.815	        &   139.7	&   3.57	&   1.97	$\pm$   0.04	&   -90.04	$\pm$   0.04	&   4.5	    $\pm$   0.1\\
13	&   2	&&  12	&   2	&   222150.010	        &   103.5	&   3.69	&   2.33	$\pm$   0.03	&   -90.07	$\pm$   0.03	&   4.84	$\pm$   0.08\\
13	&   1	&&  12	&   1	&   222162.730	        &   81.9	&   3.75	&   3.74	$\pm$   0.09	&   -90.15	$\pm$   0.06	&   5.0	    $\pm$   0.1\\
13	&   0	&&  12	&   0	&   222166.971	        &   74.6	&   3.78	&   4.43	$\pm$   0.06	&   -89.52	$\pm$   0.03	&   5.24	$\pm$   0.09\\
\hline
14	&   4	&&  13	&   4	&   239179.281      	&   201.7	&   4.34	&   0.44	$\pm$   0.02	&   -90.62	$\pm$   0.07    &   4.1	    $\pm$   0.2\\
14	&   3	&&  13	&   3	&   239211.215	        &   151.1	&   4.51	&   2.02	$\pm$   0.03	&   -90.15	$\pm$   0.03	&   4.55	$\pm$   0.07\\
14	&   2	&&  13	&   2	&   239234.034	        &   115.0	&   4.63	&   2.08	$\pm$   0.03	&   -90.09	$\pm$   0.03	&   4.51	$\pm$   0.06\\
14	&   1	&&  13	&   1	&   239247.728	        &   93.3	&   4.70	&   3.69	$\pm$   0.07	&   -90.29	$\pm$   0.05	&   5.2	    $\pm$   0.1\\
14	&   0	&&  13	&   0	&   239252.294	        &   86.1	&   4.73	&   4.25	$\pm$   0.06	&   -89.50	$\pm$   0.03	&   5.34	$\pm$   0.09\\
\hline
15	&   4	&&  14	&   4	&   256258.426$^{\ast}$	&   214.0	&   5.41	&   1.2	    $\pm$   0.1	    &   -89.1	$\pm$   0.3	    &   7.3	    $\pm$   0.7\\
15	&   3	&&  14	&   3	&   256292.630	        &   163.4	&   5.59	&   2.29	$\pm$   0.06    &   -90.14	$\pm$   0.06	&   5.1	    $\pm$   0.1\\
15	&   2	&&  14	&   2	&   256317.071	        &   127.3	&   5.72	&   2.13	$\pm$   0.06	&   -90.02	$\pm$   0.06	&   4.9	    $\pm$   0.1\\
15	&   1	&&  14	&   1	&   256331.739	        &   105.7	&   5.80	&   3.23	$\pm$   0.06	&   -90.14	$\pm$   0.04	&   4.8	    $\pm$   0.1\\
15	&   0	&&  14	&   0	&   256336.629	        &   98.4	&   5.83	&   4.00	$\pm$   0.05	&   -89.54	$\pm$   0.04	&   5.43	$\pm$   0.09\\
\hline
16	&   4	&&  15	&   4	&   273336.519	        &   227.1	&   6.64	&   0.58	$\pm$   0.02	&   -90.30	$\pm$   0.08	&   5.4	    $\pm$   0.2\\
16	&   3	&&  15	&   3	&   273372.990	        &   176.6	&   6.84	&   2.08	$\pm$   0.02    &   -90.17	$\pm$   0.03	&   4.91	$\pm$   0.07\\
16	&   2	&&  15	&   2	&   273399.051	        &   140.4	&   6.97	&   2.13	$\pm$   0.02	&   -90.16	$\pm$   0.03	&   4.93	$\pm$   0.07\\
16	&   1	&&  15	&   1	&   273414.692	        &   118.8	&   7.06	&   3.48	$\pm$   0.06	&   -90.42	$\pm$   0.04	&   5.5	    $\pm$   0.1\\
16	&   0	&&  15	&   0	&   273419.906	        &   111.5	&   7.09	&   3.82	$\pm$   0.02	&   -89.66	$\pm$   0.02	&   5.48	$\pm$   0.09\\
\hline
17	&   4	&&  16	&   4	&   290413.488$^{\ast}$	&   241.0	&   8.04	&   0.58    $\pm$   0.02    &   -90.2   $\pm$   0.1     &   5.3     $\pm$   0.2\\
17	&   3	&&  16	&   3	&   290452.224	        &   190.5	&   8.25	&   2.40	$\pm$   0.05	&   -90.47	$\pm$   0.04	&   5.4	    $\pm$   0.1\\
17	&   2	&&  16	&   2	&   290479.904	        &   154.4	&   8.40	&   2.07	$\pm$   0.03	&   -90.27	$\pm$   0.03	&   4.9	    $\pm$   0.1\\
17	&   1	&&  16	&   1	&   290496.516	        &   132.7	&   8.49	&   3.53	$\pm$   0.06	&   -90.68	$\pm$   0.04	&   5.7	    $\pm$   0.1\\
17	&   0	&&  16	&   0	&   290502.054	        &   125.5	&   8.52	&   3.81	$\pm$   0.05	&   -89.88	$\pm$   0.04	&   5.4	    $\pm$   0.1\\
\hline
18	&   4	&&  17	&   4	&   307489.264$^{\ast}$	&   255.8	&   9.62	&   0.52    $\pm$   0.01    &   -90.13  $\pm$   0.07    &   5.2     $\pm$   0.2\\
18	&   3	&&  17	&   3	&   307530.263	        &   205.3	&   9.84	&   1.88	$\pm$   0.02	&   -90.23	$\pm$   0.03	&   4.90	$\pm$   0.06\\
18	&   2	&&  17	&   2	&   307559.559	        &   169.1	&   10.00	&   1.73	$\pm$   0.02	&   -90.23	$\pm$   0.03	&   4.76	$\pm$   0.07\\
18	&   1	&&  17	&   1	&   307577.141	        &   147.5	&   10.10	&   2.91	$\pm$   0.04	&   -90.50	$\pm$   0.03	&   5.39	$\pm$   0.08\\
18	&   0	&&  17	&   0	&   307583.003	        &   140.3	&   10.10	&   3.27	$\pm$   0.05	&   -89.76	$\pm$   0.04	&   5.6	    $\pm$   0.1\\
\hline   
\end{tabularx}
\end{table*}

For each observed $J_{u}$ level, all transitions with $K$=3--0 were detected. In some cases (see Table \ref{table:CH3CCH}), lines with $K$=4 were not detectable above the 3 rms noise threshold, and were not considered in our analyses. Nevertheless, all K-ladders include a minimum of four detected transitions. The CH$_3$CCH lines present a low velocity dispersion ($\pm 0.68$ km s$^{-1}$) around the systemic velocity of the source and a mean systemic velocity of $\bar{V}_{lsr}$ = -90.0 $\pm$ 0.3 km s$^{-1}$, obtained through averaging the velocities of all 41 lines. Line profiles are narrow, with average values of 4.9 $\pm$ 0.8 km s$^{-1}$, and observed line widths also show an overall small dispersion, with FWHM values ranging from $4.15$ to $5.68$ km s$^{-1}$. This indicates that the emission originates from a quiescent region, as will be discussed in ~\S~\ref{sec:4.1}.

\subsection{Rotational diagrams and physical properties}\label{sec:3.2}

In order to estimate the excitation temperature ($T_{exc}$) and column density ($N$) of CH$_3$CCH, rotational diagrams were constructed. Under the assumption of LTE, one can derive the $T_{exc}$ and $N$ of an optically thin emission that uniformly fills the antenna beam from:

\begin{equation}
\centering
\ln\left(\frac{N_u}{g_u}\right)=\ln\left(\frac{N}{Q(T_{exc})}\right)-\frac{E_u}{k_BT_{exc}}
\label{eq:rot_diag}
\end{equation}

\noindent where Q is the species' partition function at $T_{exc}$ and k$_B$ is the Boltzmann constant. $N_u$, $g_u$ and $E_u$ are, respectively, the column density, the statistical weight and the energy of the upper level \citep{Goldsmith1999}. A plot of $\ln(N_u/g_u)$ versus $E_u$ of the observed lines will thus correspond to a straight line, whose slope and y-intercept are defined by $1/T_{exc}$ and $\ln(N/Q\mathbf{(T_{exc})})$, respectively. For a spatially unresolved emitter, a beam-dilution correction factor, $(\Delta\Omega_a/\Delta\Omega_s)$, should be introduced on the right-hand side of Equation (\ref{eq:rot_diag}). For a preliminary analysis, we adopted an average source size of 5\arcsec, in accordance with previous observations of G331 (e.g. \citealt{Hervias-Caimapo2019,Canelo2021}, and references therein).

\begin{figure}[htb]
\centering
\includegraphics[scale=0.35]{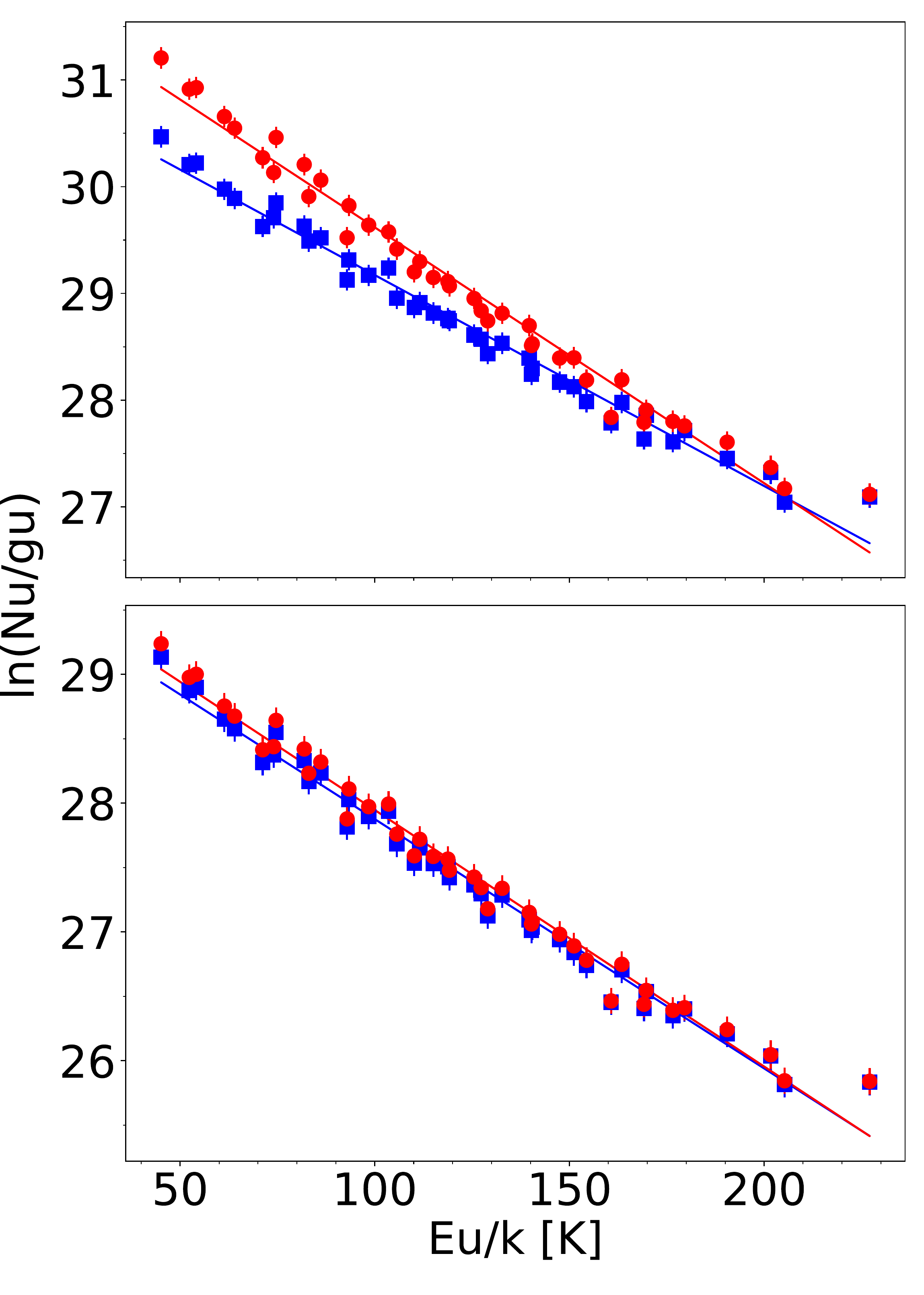}
\caption{\textit{Upper panel:} Rotational diagram of the CH$_3$CCH lines for a source size of 5\arcsec. Red circles and blue squares correspond to the lines with and without opacity corrections, respectively. Least squares fits to the points are also shown in their respective color. The error bars are propagated from the Gaussian fit to the lines and take into consideration a calibration uncertainty of 10\% \citep{Santos2021} \textit{Lower panel:} Same as the upper panel, but for a source size of 10\arcsec~ in diameter.}
\label{fig:ch3cch_rot_diag}
\end{figure}

The rotational diagram of CH$_3$CCH constructed under the assumption of optically thin lines is displayed in blue in the upper panel of Figure \ref{fig:ch3cch_rot_diag}. We derived $N$(CH$_3$CCH)~=~(2.5$\pm$0.1)~$\times$~10$^{16}$~cm$^{-2}$ and $T_{exc}$~=~50$\pm$1~K from the fit ($\chi^2_{red}$=1.49). The uncertainties were computed following the formalism of the CASSIS software\footnote{\url{http://cassis.irap.omp.eu/docs/RadiativeTransfer.pdf}}, which takes into account the calibration error of 10\%.

To explore the validity of the optically thin assumption, we estimated the lines' optical depths ($\tau$) using CASSIS. The opacity correction is incorporated to Equation (\ref{eq:rot_diag}) through iterative calculations of the optical-depth correction factor $C_{\tau}$ \citep{Goldsmith1999}, associated with the photon escape probability:

\begin{equation}
\centering
C_\tau=\frac{\tau}{1-e^{-\tau}}
\label{eq:Ctau}
\end{equation}

\noindent The optical-depth correction factor is introduced on the right-hand side of Equation (\ref{eq:rot_diag}). The iterations stop when a solution for $T_{exc}$ and $N$ converges. That is to say, when the difference between the last two iterations is smaller than 1\%. The opacity-corrected rotational diagram of CH$_3$CCH is displayed in red in both panels of Figure \ref{fig:ch3cch_rot_diag}. From the fit ($\chi^2_{red}$~=~2.57) we obtained $N$(CH$_3$CCH)~=~(4.5~$\pm$~0.3)~$\times$~10$^{16}$~cm$^{-2}$ and $T_{exc}$~=~42~$\pm$~1.0~K. This correction leads to a new scaling of the diagram's ordinate axis, resulting in a decrease of 17\% and an increase of 80\% in the values of $T_{exc}$ and $N$, respectively.

Given the rather low temperatures derived from the rotational diagrams, we infer that the CH$_3$CCH emission is originated from a cooler gas in the outer envelope of the source, in accordance with other observational works toward star-forming regions \citep{Churchwell1983,Nagy2015,Andron2018,Hervias-Caimapo2019}. Thus, we expect CH$_3$CCH to trace a more extended region. The rotational diagram constructed for an extended source (size of 10\arcsec) is displayed in the lower panel of Figure \ref{fig:ch3cch_rot_diag}. From the fit under the optically thin assumption ($\chi^2_{red}$=1.43), we derived $N$(CH$_3$CCH)=(6.9$\pm$0.4)$\times$10$^{15}$ cm$^{-2}$ and $T_{exc}$=52$\pm$1 K. 

\begin{figure*}[htb!]
\centering
\includegraphics[scale=0.35]{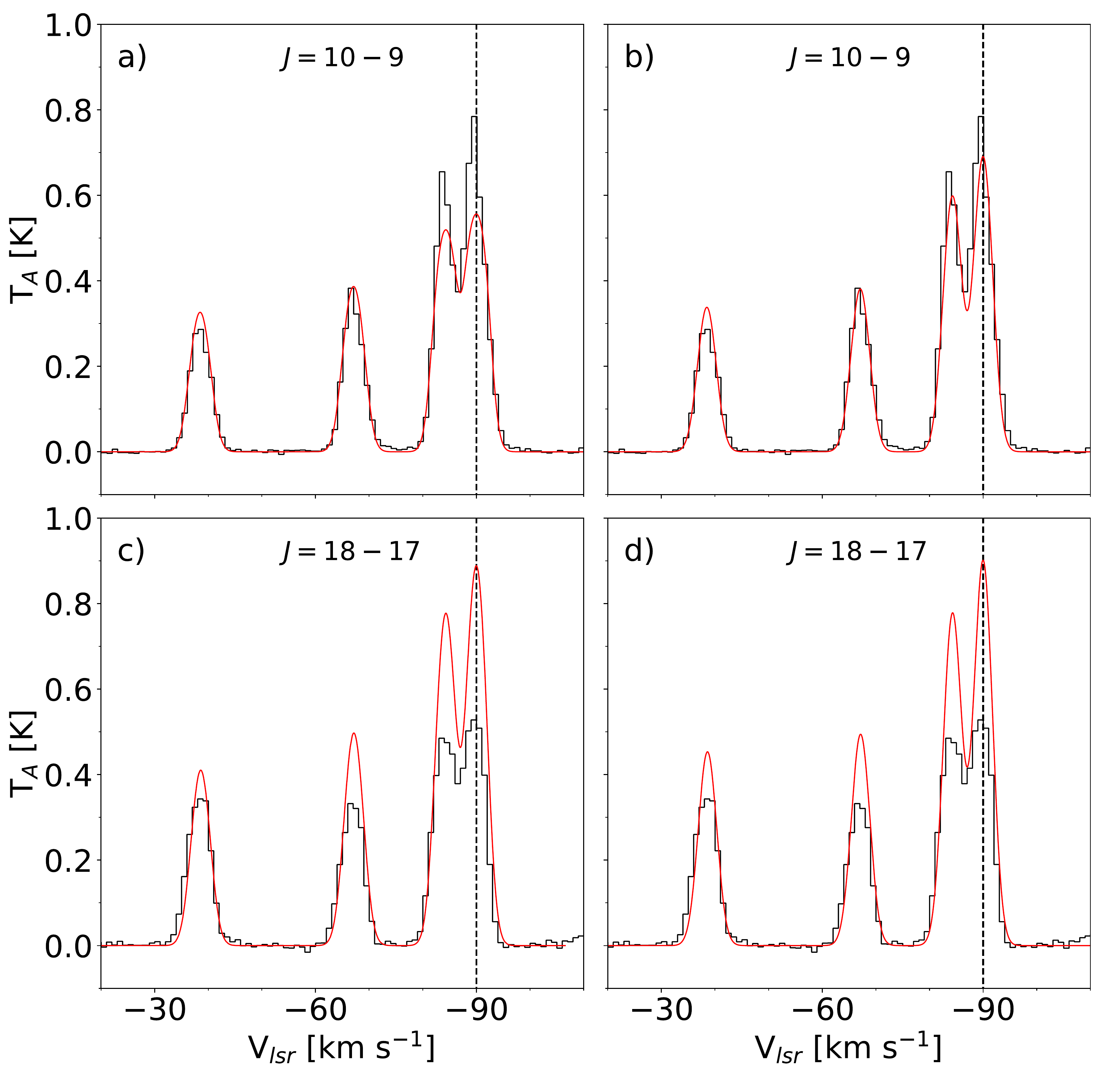}
\caption{Spectra of CH$_3$CCH observed toward G331.\textit{Upper pannels}: $J$=10--9, $\sim$~170.87--170.91~GHz; \textit{Lower pannels}: $J$=18--17, $\sim$~307.52--307.59~GHz. \textit{Left pannels}: LTE models simulated for a compact emission (5$^{\prime\prime}$). \textit{Right pannels}: LTE models simulated for an extended emission (10$^{\prime\prime}$). For all panels, the black and red spectra correspond to the observations and the LTE models, respectively. The dashed line indicates the systemic velocity of the source. The abscissas of all spectra were adjusted to the systemic velocity of the $K$=0 transition.}
\label{fig:ch3cch_lines_model}
\end{figure*}

After including the opacity correction factor, we derived $N$(CH$_3$CCH)~=~(7.5~$\pm$~0.4)~$\times$~10$^{15}$~cm$^{-2}$ and $T_{exc}$~=~50~$\pm$~1~K from the fit ($\chi^2_{red}$~=~1.51). These values were employed to model the emission of CH$_3$CCH under LTE conditions with WEEDS (see Figure \ref{fig:ch3cch_lines_model}). Overall, the synthetic spectra are in good agreement with the observations, although the accordance is sensibly higher for K-ladders with lower $J$ quantum-numbers (see the discussion in ~\S~\ref{sec:4.2}). For the 10\arcsec~source size, the contribution of C$_\tau$ to the diagram is much less significant than for more compact emissions. Indeed, the corrected fit yielded $T_{exc}$ and $N$ values only slightly different from the optically thin scenario, with respective changes of $-2.09\%$ and $8.69\%$. Table \ref{table:ntot_tex} summarizes the excitation temperatures and column densities derived from the rotational-diagram analyses.

\begin{table*}[htb!]
\caption{Values of $T_{exc}$ and $N$ of CH$_3$CCH toward G331 obtained from the analysis of the rotational diagrams.}   
\label{table:ntot_tex}      
\centering
\begin{tabular}{c c c c c}  
\hline\hline                        
Parameter                               &   \multicolumn{2}{c}{Source size: 5\arcsec}       &    \multicolumn{2}{c}{Source size: 10\arcsec}     \\
                                        &   With C$_\tau$   &   Without C$_\tau$            &   With C$_\tau$       &   Without C$_\tau$        \\
\hline
$T_{exc}$ [K]                           & 42  $\pm$   1 & 51  $\pm$   1             & 50  $\pm$   1     & 52  $\pm$   1         \\
$N$ [$\times$ 10$^{16}$ cm$^{-2}$]      & 4.5   $\pm$   0.3 & 2.5   $\pm$   0.1             & 0.75   $\pm$   0.04   & 0.69   $\pm$   0.04       \\
$\chi^2_{red}$                          &     2.57          &    1.49                       &       1.51            &   1.42                    \\
\hline  
\end{tabular}
\end{table*}

In terms of the reduced $\chi^2$ values, the fit of the 5\arcsec ~diagram with opacity corrections was less accurate in comparison with the uncorrected counterpart, whereas no significant change was observed for the 10\arcsec ~diagrams. This is a consequence of the underestimation of the source size in the 5\arcsec ~scenario, which is compensated by overestimating the lines' optical depth---in particular, for the K=0 and K=1 transitions at lower $J$ values (see Figure \ref{fig:ch3cch_lines_model}). Indeed, the CH$_3$CCH emission seems to be optically thin \citep{Churchwell1983,Fontani2002}, especially for the range of $J$ quantum-numbers observed in this work.

\subsection{Relative intensities} \label{sec:3.3}

The relative intensities of the lines within a K-ladder will change depending on the temperature of the environment. To explore this relation, the rotational spectra of CH$_3$CCH were simulated at temperatures ranging from 10 K to 100 K, using the PGOPHER general purpose software \citep{Western2016}. The rotational constants (A and B), the quartic and sextic centrifugal distortion constants (D$_J$, D$_{JK}$, H$_{J}$, H$_{JK}$ and H$_{KJ}$), and the dipole moment ($\mu$) employed in the simulations are listed in Table \ref{table:spec_params}.

\begin{table}[htb!]
\centering
\caption{Spectroscopic parameters employed in the spectrum simulations. References: $^a$JPL database; $^b$\citep{Dubrule1978}; $^c$\citep{Muenter1966}.}  
\label{table:spec_params}      
\begin{tabular}{c c| c c}  
\hline\hline
Parameter       &   Value       &   Parameter       &  Value\\
\hline
A [MHz]         &   158590$^a$      &   H$_{J}$ [Hz]    &  0.0097$^b$\\
B [MHz]         &   8545.87712$^b$  &   H$_{JK}$ [Hz]   &  0.935$^b$\\
D$_J$ [kHz]     &   2.9423$^b$      &   H$_{KJ}$ [Hz]   &  5.23$^b$\\   
D$_{JK}$ [kHz]  &   163.423$^b$     &   $\mu$ [D]       &  0.75$^c$ \\
\hline  
\end{tabular}\\
\end{table}

\begin{figure*}[htb!]
\centering
\includegraphics[scale=0.35]{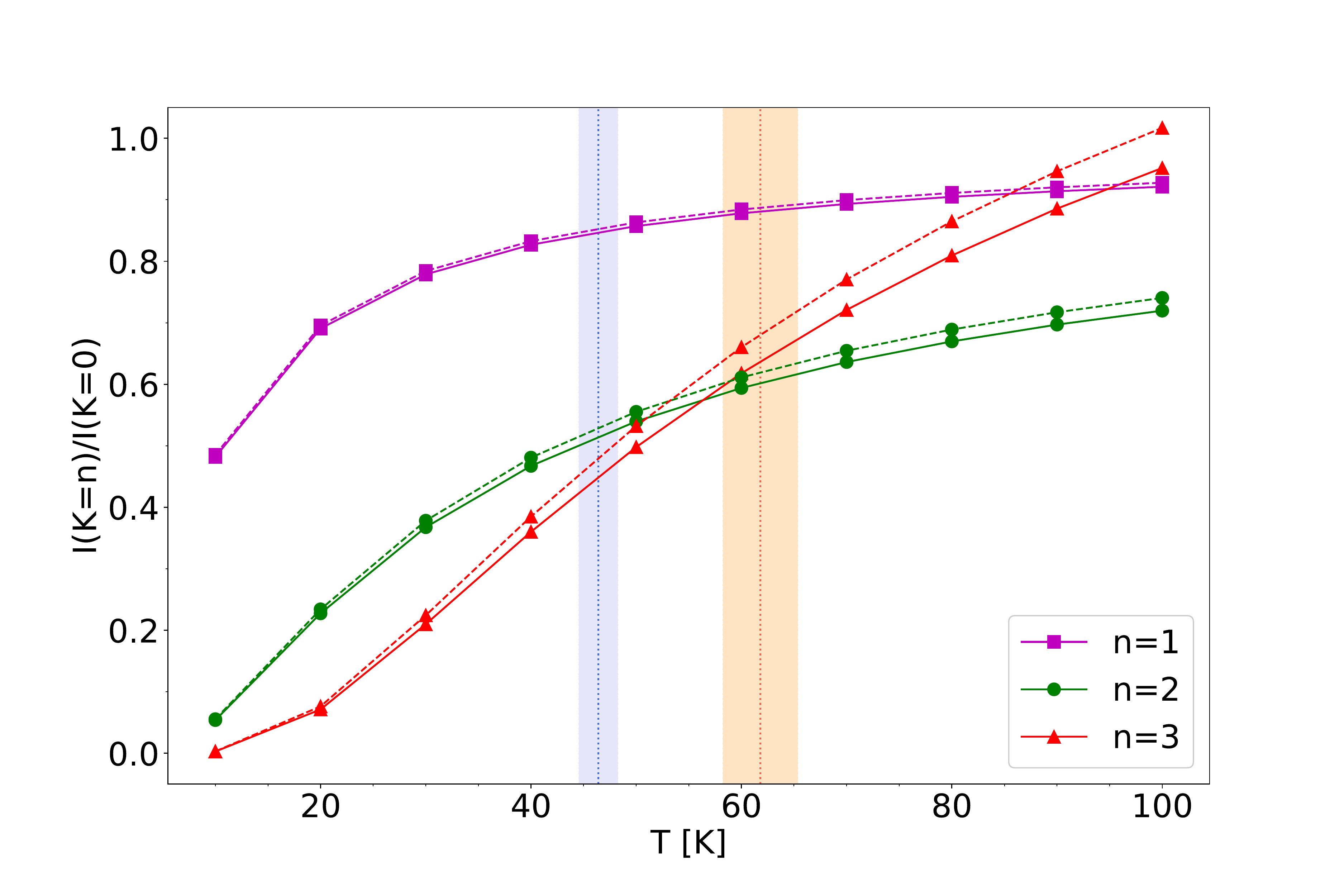}
\caption{Intensity ratios I[$K$=n]/I[$K$=0] versus temperature for transitions within a given $K$-ladder. Purple squares represent ratios with $n=1$, green circles represent ratios with $n=2$, and red triangles represent ratios with $n=3$. The solid and dashed lines correspond to, respectively, the $J$=10--9 and $J$=18--17 $K$-ladders. The red and blue regions represent upper and lower limits to the temperature profile of the source and are discussed in ~\S~\ref{sec:4.2}.}
\label{fig:ch3cch_sim_int_ratios}
\end{figure*}

The transition intensities are given by \citep{Western2016}:

\begin{equation}
\centering
I=\frac{S}{Q(T)}\left[\exp\left(\frac{-E_{l}}{k_BT}\right) - \exp\left(\frac{-E_{u}}{\mathbf{k_B}T}\right)\right],
\label{eq:intensity_spectra}
\end{equation}

\noindent which is simply the line strength $S$ times the Boltzmann factor, normalized to the partition function $Q(T)$ at the given temperature. $E_l$ and $E_u$ correspond to the energies of the lower and the upper states, respectively. In Figure \ref{fig:ch3cch_sim_int_ratios}, we plot the intensity ratios I[K=n]/I[K=0] of the predicted lines within a K-ladder versus the temperature, with $n=1$, 2 and 3. We show the results for the $J$=10--9 and $J$=18--17 transitions, corresponding to the extremes of the observed bandwidth.

As seen in the plot, the behavior of the intensity ratios are overall analogous for the range of $J$ values explored in this work. At low temperatures, the spectrum is dominated by the $K$=0 transitions, with small contributions from the other $K$ levels. As the temperature rises, however, the ratios of all $K$ transitions increase at different rates. The intensity ratios of both $K$=1 and $K$=2 transitions, compared to $K$=0, follow a similar trend. They show an initial stage of rapid growth, which becomes gradually slower as the temperature continues to rise. As can be seen for the $K$=1 transitions, the curve will eventually reach a plateau, which stabilizes the $K$=1/$K$=0 intensity ratios at values around 0.8--0.9. Comparatively, the ratios for $K$=3 present a more dramatic increase. At around 50 K, the $K$=3 lines become more intense than the $K$=2 transitions. Further, at around 100 K, their intensities surpass those of the $K$=1 transitions. At even higher temperatures, they ultimately become the most intense line of the given K-ladder. 

This behavior can be explained by nuclear spin statistics: symmetric tops with three equivalent hydrogens, such as CH$_3$CCH, have two different spin symmetries---namely, A and E. The A states correlate with the transitions with $K$=0, 3, 6, 9, ..., whereas the E states correspond to the remaining transitions, such as $K$=1, 2, 4, 5, ... (e.g. \citealt{Strom2020}). Those states have relative statistical weights of A:E = 2:1 \citep{Herzberg1945}, and therefore the population distribution of the transitions at higher temperatures will favor the $K$=3 states. At lower temperatures, however, the energy acts as a limiting factor. This phenomenon makes the CH$_3$CCH K-ladder profile highly sensitive to the local temperature, in particular with regards to the relative intensities of the K=3 transitions.

\begin{figure}[htb!]
\centering
\includegraphics[scale=0.22]{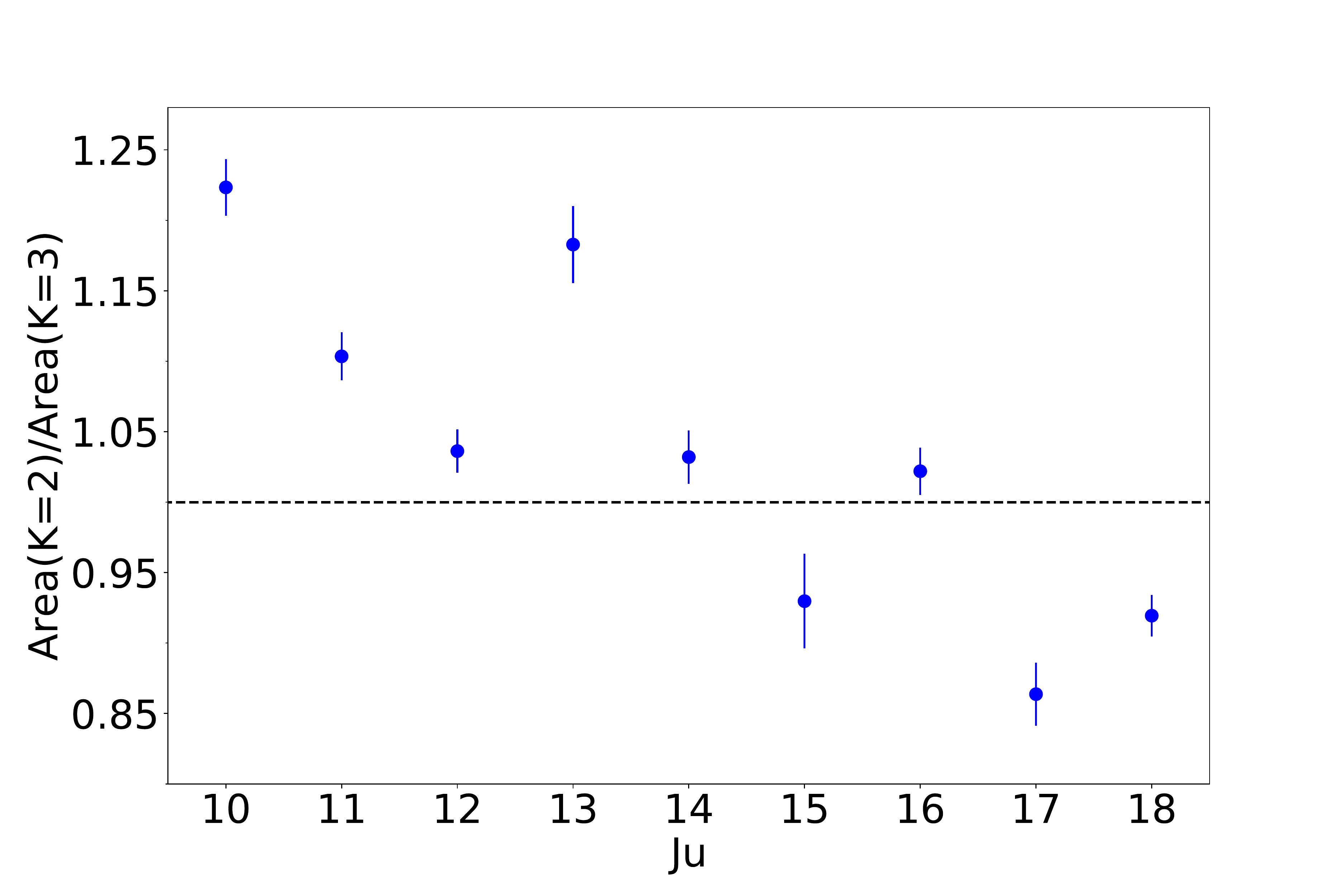}
\caption{Ratios of the areas for $K$=2 and $K$=3 lines versus the upper $J$ value of the transition. Areas and uncertainties were obtained through Gaussian fittings of the observed data using CLASS. The dashed line indicates a ratio of unity.}
\label{fig:ch3cch_area_ratio_J}
\end{figure}

As can be seen in Figure \ref{fig:ch3cch_lines}, we serendipitously observed a frequency window in which the CH$_3$CCH K-ladder profiles gradually change as a function of the rotational quantum number. For low $J$ values, which are associated with cooler regions, the $K$=2 lines are more intense than the $K$=3 (see panel a). However, as we observe transitions with higher $J$ values, which are consequently associated with warmer regions, the $K$=2/$K$=3 intensity ratios consistently decrease. From $J$=15--14 onward, the $K$=3 lines become more intense than the $K$=2 (see panel i). In Figure \ref{fig:ch3cch_area_ratio_J}, we plot the $K$=2/$K$=3 ratios of the observed areas listed in Table \ref{table:CH3CCH} as a function of the upper $J$ value. This plot yields a Pearson correlation coefficient of $r$=-0.84, which, given the small FWHM dispersion of the observed lines, clearly indicates a decreasing trend of the $K$=2/$K$=3 intensity ratios with $J$. This result strongly suggests that the lines are tracing a region with a temperature gradient, as will be discussed in ~\S~\ref{sec:4.2}.

\subsection{Search for the CH$_3$CCH isotopologues}\label{sec:3.4}

The $^{13}$C isotopologues of CH$_3$CCH, as well as its deuterated forms, have been successfully detected toward multiple interstellar sources \citep{Gerin1992,Markwick2002,Markwick2005,Belloche2013,Potapov2016,Halfen2017,Schmidt2019,Agundez2019}. Recently, \cite{Agundez2021} have detected the two doubly deuterated forms of methyl acetylene---CHD$_2$CCH and CH$_2$DCCD---toward the dense core L483. They derived abundance ratios of CH$_3$CCH/CHD$_2$CCH$=$34$\pm$10 and CH$_3$CCH/CH$_2$DCCD$=$42$\pm$13, which are only a few times less than the singly deuterated counterparts. 

\begin{figure}[htb]
\centering
\includegraphics[scale=0.25]{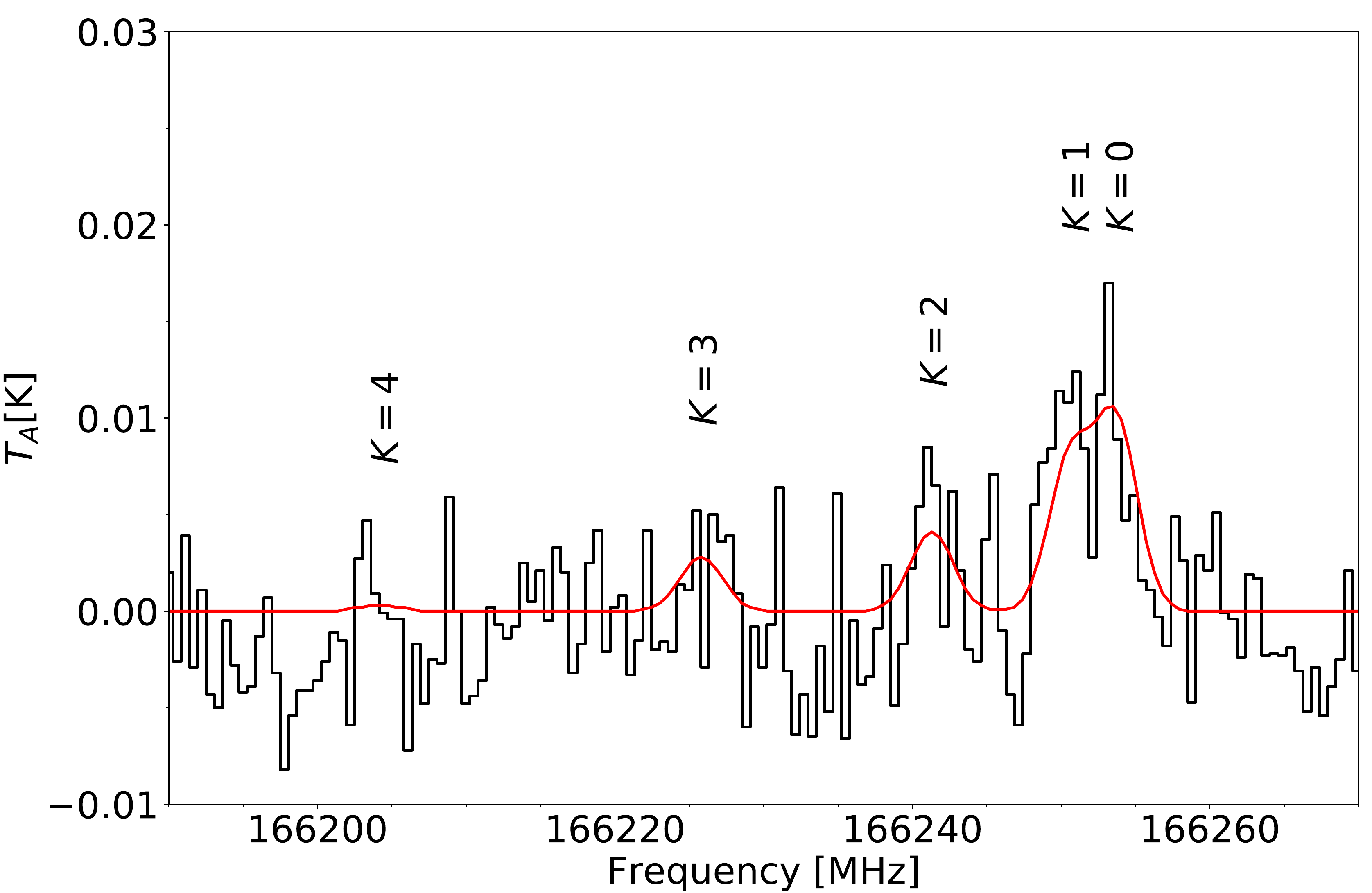}
\caption{Modelled 10$_K$--9$_K$ K-ladder of $^{13}$CH$_3$CCH (in red) superimposed on the spectrum observed toward G331 (in black).}
\label{fig:13ch3cch}
\end{figure}

Motivated by the copious amount of CH$_3$CCH line detections in G331, we searched across the survey for the $^{13}$C and D isotopologues of CH$_3$CCH. Only the 10$_0$--9$_0$ and 10$_1$--9$_1$ lines of $^{13}$CH$_3$CCH were detected above the 3$\sigma$ threshold. However, they are blended, and just marginally above the detection limit, and so we cannot reliably confirm these detections. Assuming  a ratio of  $^{12}$C/$^{13}$C~$\sim$~20, as estimated for G331 and other sources toward the Galactic center \citep{Wilson1994,Requena-Torres2006,Mendoza2018,Duronea2019,Yan2019}, we derive a $^{13}$CH$_3$CCH column density of $N\sim3.75\times10^{14}$ cm$^{-2}$. This column density was used to model the emission of the 10$_K$--9$_K$ K-ladder of $^{13}$CH$_3$CCH under LTE conditions (Figure \ref{fig:13ch3cch}), which is shown to be consistent with the observed spectrum of G331.

\section{Discussion} \label{sec:discussion}

\subsection{Origin of the emission}\label{sec:4.1}

\begin{figure}
\centering
\includegraphics[scale=0.35]{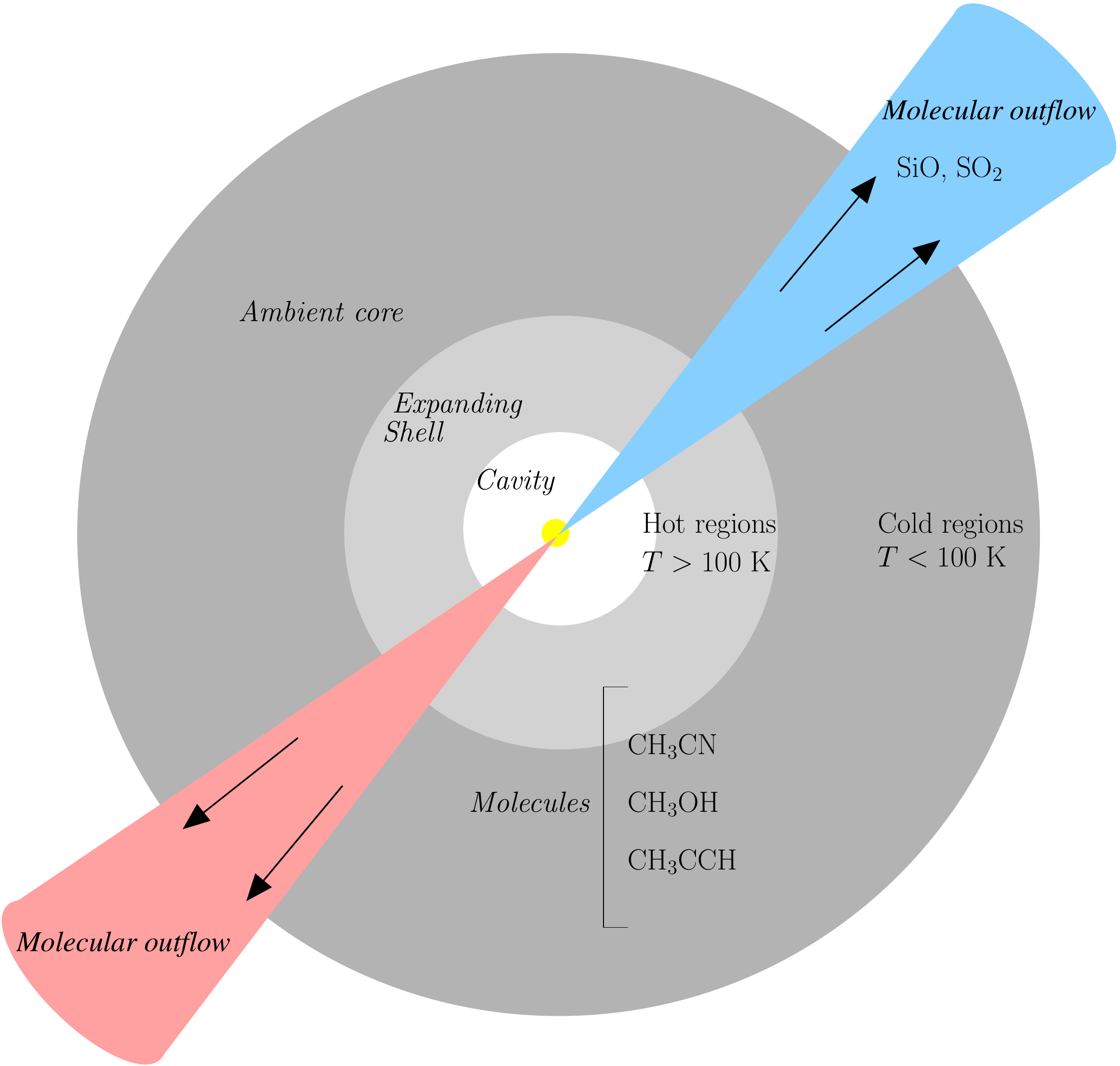}
\caption{Schematic view of the physical model proposed for G331. Based on the models of \cite{Merello2013,Duronea2019,Hervias-Caimapo2019}.}
\label{fig:G331_MODEL}
\end{figure}

\citet{Hervias-Caimapo2019} have observed several molecular lines toward G331 and classified them according to the line profile, with narrow lines tracing the emission from the core ambient medium and lines with broad wings tracing the outflow and shocked region. Accordingly, we expect from the narrow line profiles that the observed CH$_3$CCH emission is originated from a quiescent (not expanding) core medium. This region is also associated with lower temperatures (below 100 K), which is consistent with the excitation temperatures derived from the rotational diagram. A schematic view of the physical model of G331, which considers the emission of CH$_3$CCH and other molecular tracers, is presented in Figure \ref{fig:G331_MODEL}.

\subsection{Physical conditions: gas temperatures}\label{sec:4.2}

In ~\S~\ref{sec:3.4}, we discussed the dependence of the $K$=2/$K$=3 intensity ratios of CH$_3$CCH with increasing temperatures. This inversion in relative populations within the same K-ladder is observed toward G331, and can be easily inferred by visual inspection (see Figure \ref{fig:ch3cch_lines}).

\begin{figure}[htb!]
\centering
\includegraphics[scale=0.22]{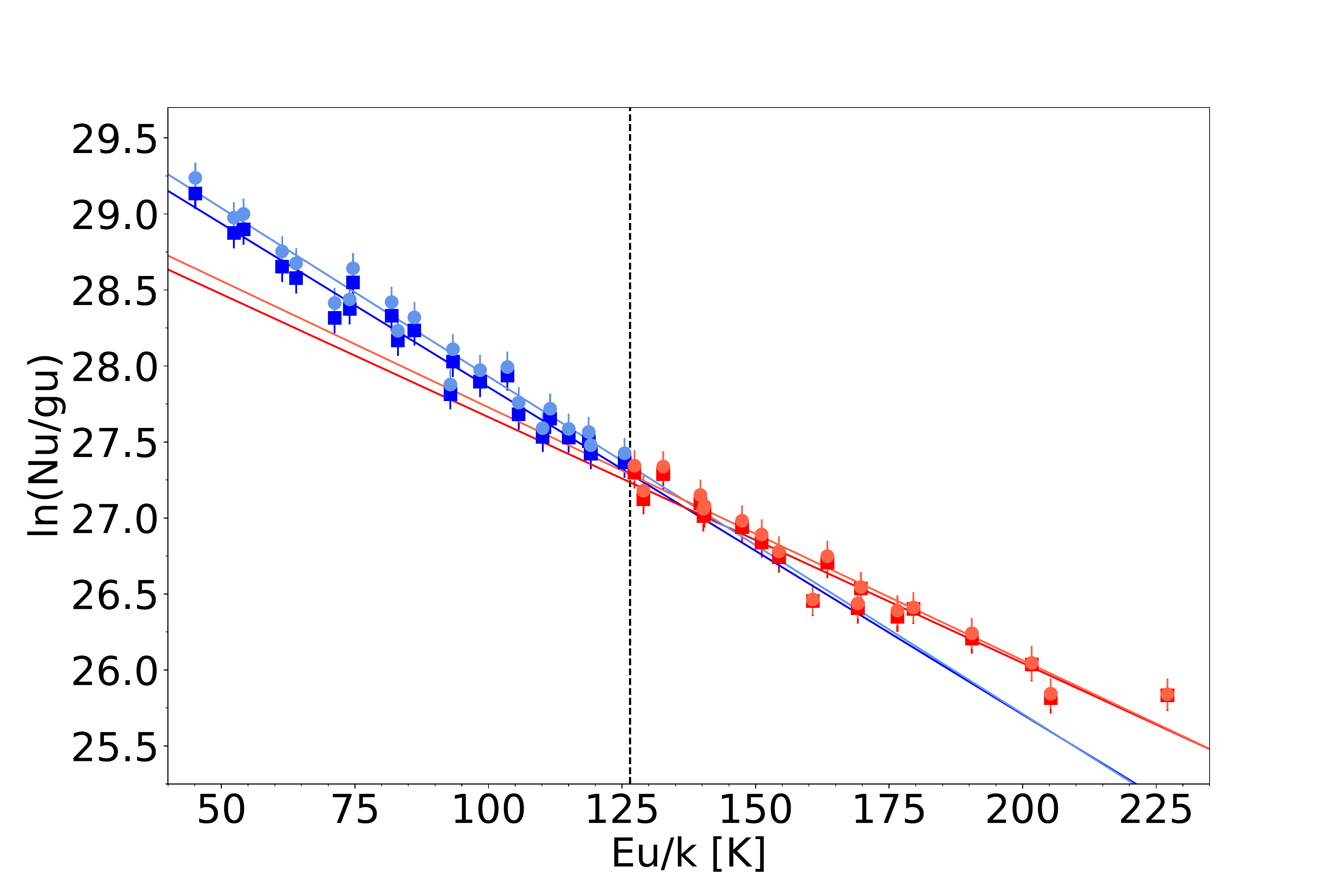}
\caption{Rotational diagram of the CH$_3$CCH lines for a source size of 10\arcsec. The points were divided into two subsets indicated by the dashed line. The points pertaining to the low- and high-energy subsets are shown in blue and red, respectively. Least squares fits to the points within each subset are shown. Circles and squares correspond to the lines with and without opacity corrections, respectively.}
\label{fig:rot_diag_2_comp}
\end{figure}

States with higher $J$ quantum numbers require higher energies to be populated, and therefore tend to be associated with warmer regions than the lower $J$ states. Given that the $K$=2/$K$=3 intensity ratios are negatively correlated with $J$ (see Figure \ref{fig:ch3cch_area_ratio_J}), we suggest that the CH$_3$CCH emission is in fact described by a temperature gradient, which explains the behavior seen in Figure \ref{fig:ch3cch_lines}. In order to estimate the extent of the temperature gradient, we have grouped the points in the rotational diagram into two subsets (Figure \ref{fig:rot_diag_2_comp}) separated at $E_{up}$ $\approx$125 K---according to the behavior of the $K$=2/$K$=3 intensity ratio of the K-ladder: the lower-energy group contains all $K$=2 transitions with $J\leq$~14, for which K=2/K=3~$\gtrsim$~1. Contrarily, the higher-energy subset encompasses all $K$=2 transitions with $J>$~14, for which K=2/K=3~$\lesssim$~1. The resulting $T_{exc}$ and $N$ obtained from the linear fit of each subset are listed in Table \ref{table:ntot_tex_2_comp}.

\begin{table*}[htb!]
\caption{Values of $T_{exc}$ and $N$ of CH$_3$CCH obtained from the 2-component rotational diagrams for a source size of 10\arcsec.}   
\label{table:ntot_tex_2_comp}      
\centering
\begin{tabular}{c c c c c}  
\hline\hline                        
Parameter                               &   \multicolumn{2}{c}{Lower-energy comp. (blue)}       &    \multicolumn{2}{c}{Higher-energy comp. (red)}  \\
                                        &   With C$_\tau$   &   Without C$_\tau$                    &   With C$_\tau$       &   Without C$_\tau$            \\
\hline
$T_{exc}$ [K]                           & 45  $\pm$   2 & 46  $\pm$   2                     & 60  $\pm$   3     & 62 $\pm$   3         \\
$N$ [$\times$ 10$^{15}$ cm$^{-2}$]      & 7.9   $\pm$ 0.6   & 7.2   $\pm$   0.5                     & 5.7   $\pm$   0.9    & 5.3   $\pm$   0.8       \\
$\chi^2_{red}$                          &     0.87          &    0.82                               &       1.19            &   1.14                    \\
\hline  
\end{tabular}
\end{table*}

Assuming an optically thin scenario, the rotational diagram divided into two subsets yields temperature components of $\sim$45 and $\sim$60 K. Although the points in the rotational diagram show an overall fairly linear trend, the $T_{exc}$ obtained from this analysis is highly sensitive to the fitted slope. Consequently, a difference of around 15 K arises between the derived values from the linear fittings of the two subsets. This temperature interval coincides with the region of Figure \ref{fig:ch3cch_sim_int_ratios} where the intensity inversion of $K$=2 and $K$=3 occurs. Thus, one can regard the two temperature components extracted from the blue and red subsets in Figure \ref{fig:rot_diag_2_comp} as, respectively, lower and upper limits to the temperature profile of the observed emitting region---indicated as the blue and red regions in Figure \ref{fig:ch3cch_sim_int_ratios}. The rotational diagram analysis of \S~\ref{sec:3.2} is therefore suited to evaluate the large-scale emission of CH$_3$CCH, yielding an averaged temperature of $T_{exc}\sim$ 50.2 K. The intrinsic spectroscopic properties of methyl acetylene, however, enabled us to perform a more direct and meticulous assessment of the temperature of the source. The same discussion is valid for the optically thick scenario.

The synthetic spectra under the LTE assumption shown in Figure \ref{fig:ch3cch_lines_model} reproduces more accurately the observed data for lower $J$ quantum-numbers. This gradual loss in accuracy for higher $J$ transitions is likely a consequence of the temperature profile that leads to the intensity inversion around $J_u$=14 (see Figure \ref{fig:ch3cch_area_ratio_J}). Indeed, throughout the entire $\sim$170840--307600 MHz frequency range, modeled transitions with $K$=3 are less intense than the $K$=2 counterparts. Since using only one temperature component is not completely adequate to describe our observations, the model will be increasingly amiss as the bulk of the emission transits to regions with different average gas temperatures. Hence, our LTE model seems to best describe the cooler environment, traced by the lines with lower $J$ values.

Previously, \citet{Hervias-Caimapo2019} have used ALMA to observed four CH$_3$CCH lines within the $J$=21--20 K-ladder toward G331, which showed a peak of emission located at a radius of $\sim$1.2\arcsec. From the rotational diagram constructed with the CH$_3$CCH lines, they obtained an excitation temperature of $T_{exc} = 70\pm7$ K. Their results are in accordance with our hypothesis of a temperature gradient, which indeed predicts that transitions with higher $J$ quantum-numbers will be associated with higher excitation temperatures and more compact regions.

\subsection{Kinematics}\label{sec:4.3}

\begin{figure*}[htb!]\centering
\includegraphics[scale=0.35]{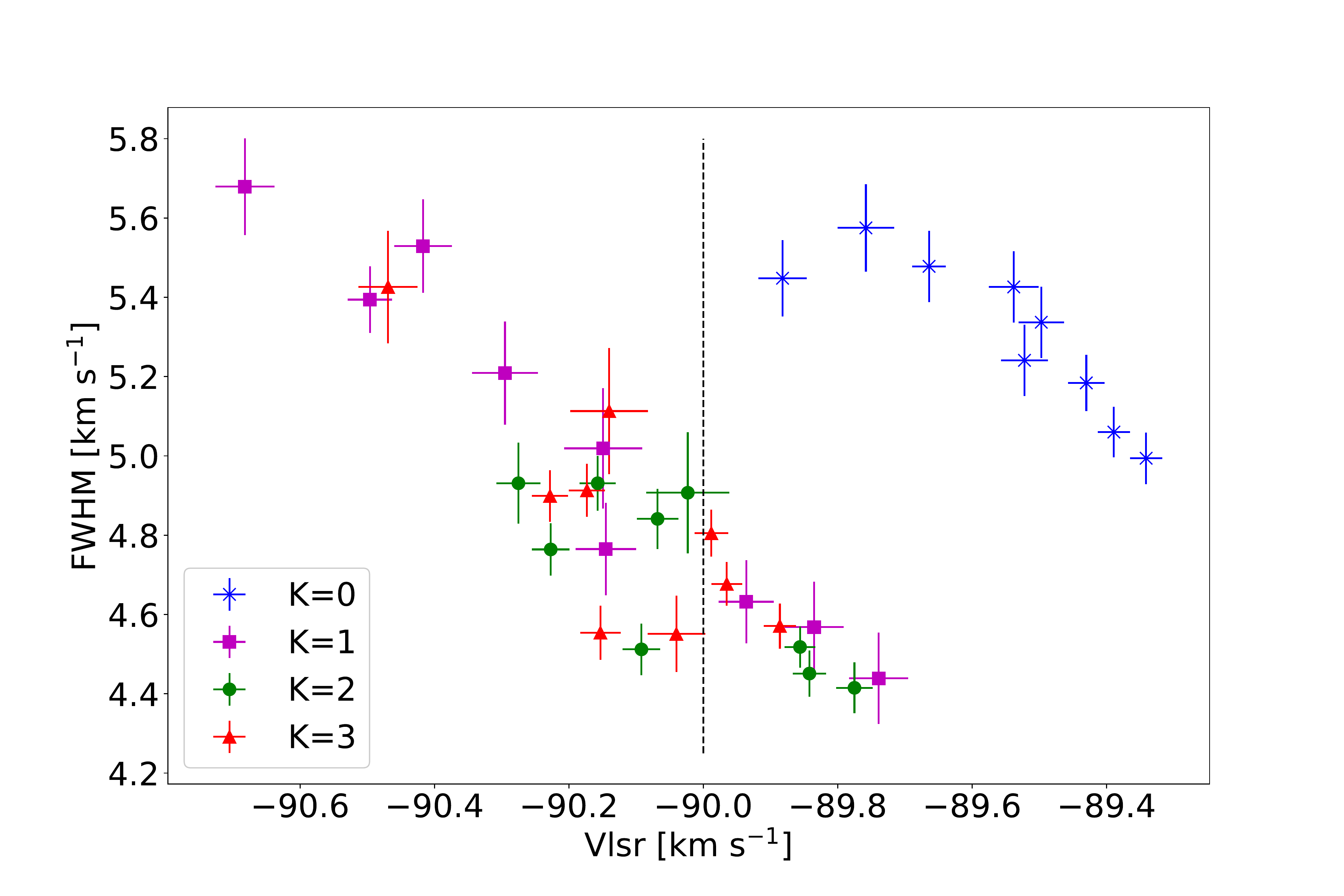}
\caption{Line widths vs. peak velocities obtained from Gaussian fittings of the observed transitions. Blue crosses represent transitions with $K$=0, purple squares represent transitions with $K$=1, green circles represent transitions with $K$=2 and red triangles represent transitions with $K$=3. The black dashed line separates the blue- and redshifted regions relative to the adopted systemic velocity of G331 (-90 km s$^{-1}$).
}
\label{fig:fwhm_x_vlsr}
\end{figure*}

The spectroscopic parameters (see Table \ref{table:CH3CCH}), obtained through fitting a Gaussian profile to each line, are directly related to the kinematics of the emitting region---or, at least, of the gas where the bulk of the emission is originated. In Figure \ref{fig:fwhm_x_vlsr}, we present a plot of the widths and peak velocities of the observed lines, which are listed in Table \ref{table:CH3CCH}.

The line widths and peak velocities of transitions with the same quantum number $K$ are overall strongly correlated. Indeed, the Pearson correlation coefficient of the $K$=0, $K$=1, $K$=2 and $K$=3 groups are, respectively, $r$=-0.85, $r$=-0.97, $r$=-0.78, and $r$=-0.80. This coefficient measures the linear correlation between two variants by means of the ratio of their covariance and the product of their standard deviation. Thus, $r$=$\pm$1 correspond to perfect linear correlations (either positive or negative) and $r$=0 corresponds to no correlation. The lines with $K$=4 are fainter and harder to observe, and therefore were not included in the comparison. Their smaller signal-to-noise ratios hampers their analysis, which is reflected in their higher associated uncertainties reported in Table \ref{table:CH3CCH}. Nonetheless, for each $K$ group, there is a clear tendency for broader lines to present increasingly blueshifted velocities. The same trend was also observed by \citet{Difrancesco2004} for bright ($T_B^{\text{max}} \geq 10 ~\sigma$) N$_2$H$^+$ lines toward the Ophiuchus A star-forming core: they found that the higher velocity lines were systematically broader, which they associate to infalling motions that increase local rotation speeds.

Interestingly, the $K$=0 lines are systematically broader and less blueshifted than the other groups, indicating that they might be tracing a slightly different region than the bulk of the emission of the other $K$ transitions. At low temperatures, the contributions from transitions with $K\neq0$ to the rotational spectrum of CH$_3$CCH are minor (see Figure \ref{fig:ch3cch_sim_int_ratios}), and the emission is dominated by the $K$=0 lines. As the temperature rises, the relative intensities of the $K\neq0$ transitions rapidly increase, resulting in richer K-ladders. Thus, it is likely that the bulk of the $K$=0 emission is originated from a cooler and more extended region, which does not contribute as significantly to the lines with $K\neq0$. These, in turn, must be tracing a hotter and more compact environment, resulting in a different kinematic signature. The systematic redshift of the peak velocities for the $K$=0 lines compared to the other groups also suggests interesting velocity features of the cold component. It is worth mentioning that the mean velocity shift between lines with the same quantum number $K$ within the $J$=10--9 and $J$=18--17 K-ladders is $\sim$0.49 km s$^{-1}$, which is more than three times the poorest obtained velocity resolution. Further surveys of the CH$_3$CCH emission toward G331 with higher angular resolution are imperative to better understand the different gas components, and the small-scale structure of the hot core.

\begin{figure}[htb!]\centering
\includegraphics[scale=0.25]{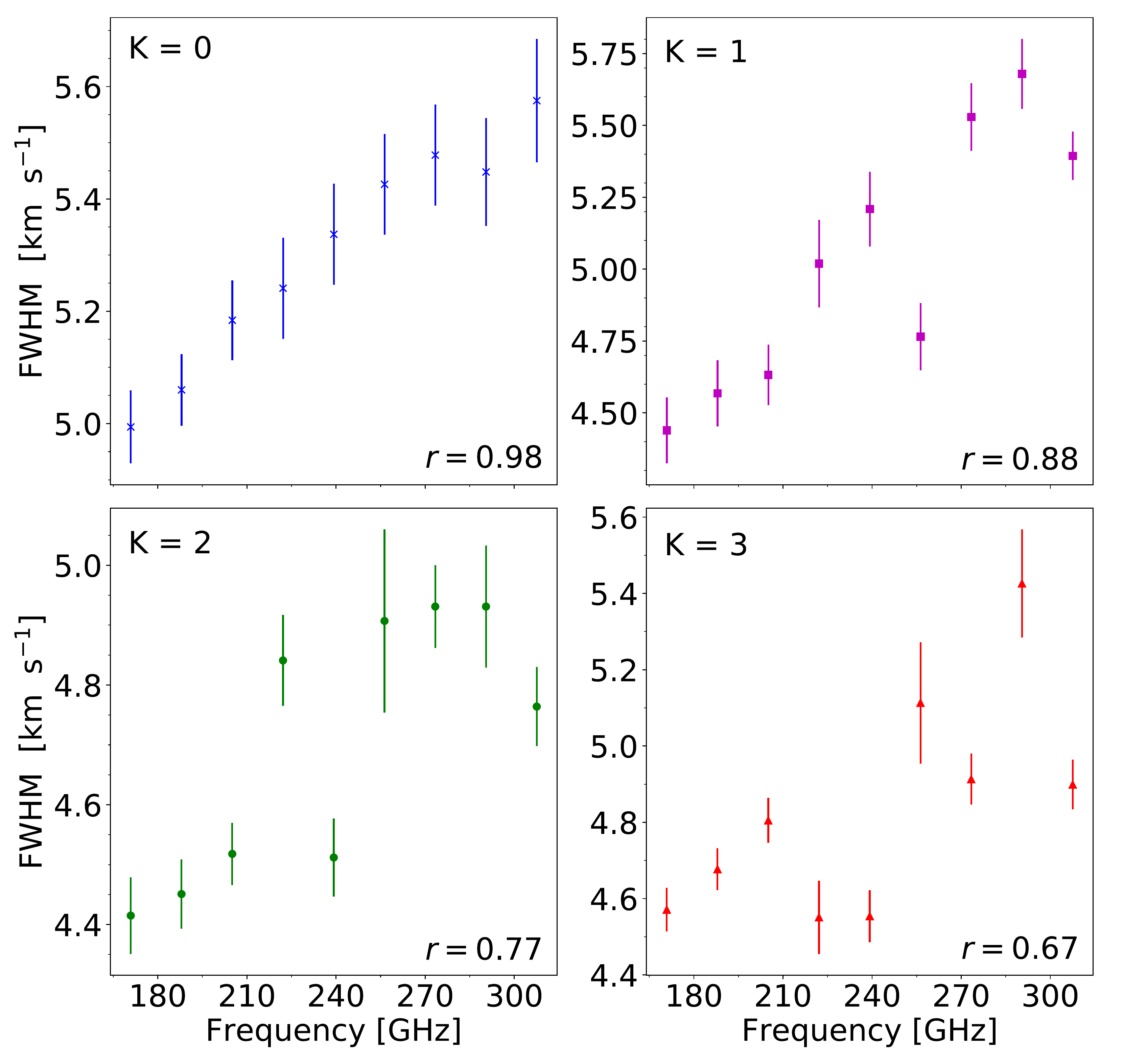}
\caption{Line widths vs. rest frequencies obtained from Gaussian fittings of the observed transitions with $K$=0, $K$=1, $K$=2 and $K$=3.}
\label{fig:fwhm_x_freq}
\end{figure}

In Figure \ref{fig:fwhm_x_freq}, we show a comparison of the line widths versus the rest frequency, together with their Pearson coefficients. From the plots, it is clear that the FWHM is correlated to the line frequency, particularly for the lines with $K$=0 and $K$=1. Since thermal broadening effects are negligible for radio transitions at the considered temperature range (see, for example, the discussion of \citealt{Fontani2002}), the broader widths associated with higher $J$ values indicate that the gas where the bulk of the emission is originated must become gradually more turbulent as the temperature increases. This is fairly reasonable, considering that the warmer emission comes from the inner parts of the envelope surrounding the Massive Young Stellar Object (MYSO).

\subsection{CH$_3$CCH abundances}\label{sec:4.4}

The abundance of CH$_3$CCH relative to H$_2$ ($X$(CH$_3$CCH)) is defined as the ratio of the column densities $N$(CH$_3$CCH)/$N$(H$_2$). We have estimated the H$_2$ column density in G331 from ALMA measurements of N(H$^{13}$CO$^+$)$\approx$(1.5--3.2)$\times$10$^{13}$~cm$^{-2}$ \citep{Merello2013b, Merello2013, Hervias-Caimapo2019}, adopting an abundance ratio $N$(H$^{13}$CO$^+$)/$N$(H$_2$)=3.3$\times$10$^{-11}$, as measured by \cite{Blake1987} in Orion KL. Additionally, a hydrogen column density of $N$(H$_2$)=2.7$\times$10$^{23}$~cm$^{-2}$ was reported for G331 from the continuum emission's surface density at 1.2 mm \citep{Duronea2019}. Considering those two limits for $N$(H$_2$), we derived a ratio of $X$(CH$_3$CCH)$\approx$(0.8--2.8)$\times10^{-8}$. 

In Table \ref{tab:abundances_ch3cch}, we present a comparison of the CH$_3$CCH abundances with respect to H$_2$ in different massive star-forming regions, including G331. Our derived abundance and excitation temperature agree well with the ones obtained for similar sources, which indicates similarities in the formation pathway of this molecule at such environments. The fact that this abundance comes mostly from the enveloping gas reinforces the conclusions drawn by other works (e.g., \citealt{Oberg2013,Fayolle2015,Giannetti2017,Taniguchi18}) that the formation of methyl acetylene is not limited by heat input, and should take place during the earlier stages of star-formation. In other words, CH$_3$CCH is likely a so-called zeroth-generation molecule \citep{Herbst2009}. Nonetheless, in the case of solid-phase formation, heat will eventually be required in order for the molecule to desorb to the gas-phase.

\begin{table*}[ht]
    \centering
       \caption{Abundances of CH$_3$CCH relative to H$_2$ and CH$_3$OH observed toward high-mass star-forming sites. References: $^a$\citep{Taniguchi18}, $^b$\citep{Giannetti2017}, $^c$\citep{Fayolle2015}. $^\ast$Derived from the average abundances reported by \cite{Giannetti2017}. n.p. = not provided. The excitation temperatures are given in units of K.}
    \begin{tabular}{cccc}
\hline\hline
Source              &   /$N$(H$_2$)                         &   /$N$(CH$_3$OH)              &   T$_{exc}^{\text{CH$_3$CCH}}$\\
\hline
G12.89+0.49$^a$     &   4.2$^{+2.8}_{-2.2} \times 10^{-8}$  &   0.34$^{+0.28}_{-0.15}$      &   33$^{+20}_{-9}$\\ 
G16.86−2.16$^a$     &   3.2$^{+1.9}_{-1.5} \times 10^{-8}$  &   0.36$^{+0.28}_{-0.18}$      &   29$^{+15}_{-8}$\\   
G28.28−0.36$^a$     &   7.6$^{+5.2}_{-4.4} \times 10^{-8}$  &   1.61$^{+1.4}_{-0.86}$       &   23$^{+9}_{-6}$\\
ATLASGAL$^b$        &   (0.5--2.5)$\times10^{-8}$~$^b$      &   0.31$^\ast$                 &   34.5$^{+24.6}_{-10.5}$\\
NGC 7538 IRS9$^c$   &            n.p.                       &   1.3 $\pm$ 0.4               &   47 $\pm$ 5\\
W3 IRS5$^c$         &           n.p.                        &   2.2 $\pm$ 0.7               &   58 $\pm$ 8\\
AFGL490$^c$         &           n.p.                        &   1.8 $\pm$ 0.8               &   41 $\pm$ 7\\
G331 (this work)    &   (0.8--2.8)$\times10^{-8}$           &   0.42 $\pm$0.05              &   50 $\pm$ 1\\
\hline
    \end{tabular}
    \label{tab:abundances_ch3cch}
\end{table*}

We also estimate the fractional abundance of CH$_3$CCH with respect to methanol, $N$(CH$_3$CCH)/$N$(CH$_3$OH), to be 0.42 $\pm$ 0.05. This value is listed in Table \ref{tab:abundances_ch3cch}, together with the ratios obtained from the literature for other massive star-forming regions. The methanol column density in G331 is assumed to be of $\approx(1.8\pm0.2)\times10^{16}$ cm$^{-2}$, as derived by \cite{Mendoza2018} under the LTE formalism and assuming the total CH$_3$OH abundance as the sum of the A-CH$_3$OH and E-CH$_3$OH contributions. This methanol emission, however, is thought to originate from a region of only 5\arcsec.3, which is considerably more compact than the emitting region of CH$_3$CCH. Given that the envelope abundance of methanol in G331 is not fully known, the derived $N$(CH$_3$CCH)/$N$(CH$_3$OH) ratio is likely artificially lower than in reality. Nonetheless, it is still fairly close to the ratios derived in other sources, further strengthening the conclusion that they present similar chemistry.

\section{Chemical modeling} \label{sec:modelling}

Formation pathways of CH$_3$CCH in the ISM include gas-phase ion-neutral routes with C$_2$H$_2^+$ as a precursor \citep{Schiff1979,millar1984}, neutral-neutral reactions such as CH $+$ C$_2$H$_4$ $\to$ CH$_3$CCH $+$ H \citep{Turner1999}, as well as dissociative recombination reactions involving larger hydrocarbons \citep{Calcutt2019}. Grain-surface reactions are also proposed in order to explain the observed abundances of CH$_3$CCH \citep{Hickson2016,Guzman2018}. Regarding massive environments, \cite{Taniguchi2019} have constructed hot-core models to investigate the formation pathways of cyanopolyynes and other carbon-chain species, including CH$_3$CCH and c-C$_3$H$_2$, around MYSOs. They found chemical similarities between  methyl acetylene, methane and cyanopolyynes, being all triggered by CH$_4$ sublimation from dust grains. Furthermore, those species are also shown to accumulate in the bulk of the ice until the temperature reaches their respective sublimation points. Comparatively, small reactive hydrocarbons such as CCH and CCS are shown not to accumulate onto dust grains, being readily destroyed in the gas-phase instead. Complementary, \cite{Andron2018} discuss the formation routes of methyl acetylene in the solar-type protostar IRAS 16293-2422. They show that, at large distances form the central star, the abundance of CH$_3$CCH in the solid phase is higher than in the gas-phase, which indicates that it is efficiently produced on dust grains at low temperatures through successive hydrogenation of C$_3$. Moreover, their models also indicates that reactions in the gas-phase with C$_2$H$_4$ and C$_3$H$_5$ as precursors can also effectively form methyl acetylene in the outer envelope.

In order to investigate the chemical evolution of CH$_3$CCH in G331, we constructed a time-dependent chemical model of the source using the three-phase gas-grain Nautilus code \citep{Ruaud2016}, in which the grain surface chemistry is distinguished from the bulk of the ice. The model consists of two zero-dimensional steps (i.e., the physical properties of the source are uniform and static), comprising the dark-cloud and the hot-core phases of star formation. The 1$^{st}$ step, representing an initial dark-cloud phase, was computed using the elemental abundances listed in Table \ref{table:init_naut}, as was computed in previous works about the chemistry of HNCO in G331 \citep{Canelo2021}. For the physical parameters, we adopted  a gas temperature and density of 10~K and 1 $\times$ 10$^4$~cm$^{-3}$, respectively. Then, we employed the abundances derived from the dark-cloud model after 10$^5$~yr as initial parameters for the 2$^{nd}$ step: a \lq\lq rapid\rq\rq \ hot-core phase. For this second period we used $T=80$~K and tested four different values for $n_{H_2}$: $1\times10^6$, $5\times10^6$, $1\times10^7$ and $5\times10^7$ cm$^{-3}$. The gas and dust temperatures were set as equal throughout the entire simulation, as they have been shown to be coupled in dense sources \citep{Merello2019}. Moreover, standard visual extinction ($A_V=10$ mag) and cosmic ray ionization rates ($\zeta$=1.3$\times$10$^{-17}$~cm$^{-3}$) were used. The standard chemical network presented in the KInetic Database for Astrochemistry (KIDA)\footnote{\url{http://kida.obs.u- bordeaux1.fr/}} \citep{Wakelam2015} catalog was employed in the simulations.

\begin{table}[htb!]
\centering
\caption{Initial gas-phase elemental abundances, with the format a(b) representing a$\times$10$^b$. References: 1. \citet{Wakelam2008}; 2. \citet{Jenkins2009}; 3. \citet{Hincelin2011}; 4. low-metal abundances from \citet{Graedel1982}; 5. depleted value from \citet{Neufeld2005}.}   
\label{table:init_naut}      
\begin{tabular}{l c c l c c}  
\hline\hline
Element &   n$_i$/n$_H^a$   &   Ref.    &   Element &   n$_i$/n$_H^a$   &   Ref.\\
\hline
H$_2$   &   0.5             &           &   He      &   9.0(-2)         &   1\\
N       &   6.2(-5)         &   2       &   O       &   2.4(-4)         &   3\\
C$^+$   &   1.7(-4)         &   2       &   S$^+$   &   1.5(-5)         &   2\\
Fe$^+$  &   3.0(-9)         &   4       &   Si$^+$  &   8.0(-9)         &   4\\
Na$^+$  &   2.0(-9)         &   4       &   Mg$^+$  &   7.0(-9)         &   4\\
Cl$^+$  &   1.0(-9)         &   4       &   P$^+$   &   2.0(-10)        &   4\\
F       &   6.7(-9)         &   5       &           &                   &   \\
\hline  
\end{tabular}
\end{table}

\begin{figure}[htb!]\centering
\includegraphics[scale=0.237]{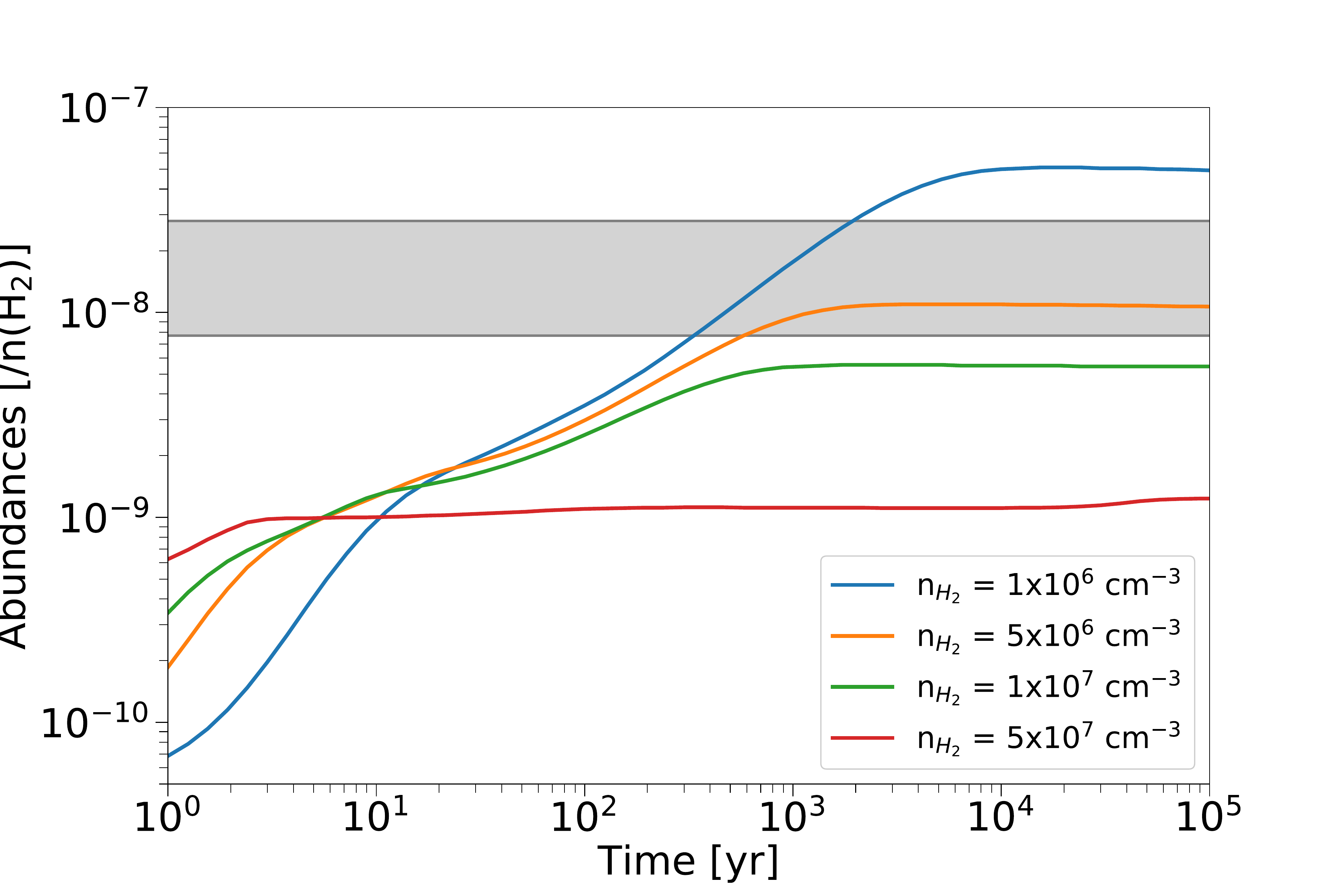}
\caption{Abundance of CH$_3$CCH as a function of time, obtained from the hot-core model for different initial densities of H$_2$. The gray area indicates the methyl-acetylene abundance derived from the observations.}
\label{fig:model_ch3cch}
\end{figure}

The temporal-evolution of the CH$_3$CCH abundance derived from the hot-core model for the different hydrogen densitites is presented in Figure \ref{fig:model_ch3cch}. The abundance derived from our observations is best predicted by the model with $n_{H_2}=5\times10^6$ at timescales as early as $\sim$~10$^3$~yr, in agreement with the expected age of this source \citep{Merello2013b}. This is consistent with an extended and thus less dense emitting gas, as was inferred from the radiative analyses. Furthermore, the steady-state plateau in abundance reached after $\sim$~10$^3$~yr is consistent with the hypothesis of CH$_3$CCH accumulating onto dust grains and enriching the gas-phase upon desorption. Experiments on interstellar ice analogues exposed to energetic electrons ensued the formation of CH$_3$CCH, among other related molecules \citep{Abplanalp2019}, which supports this hypothesis. Complementary observations of different species with related chemical networks are desirable to build a comprehensive view on the properties of the source, and will be included in future works focused on constructing a thorough chemical and physical model of G331.

\section{Conclusions} \label{sec:conclusions}

We have conducted a spectral survey of CH$_3$CCH toward the Hot Molecular Core G331.512–0.103, resulting in the detection of 41 lines without contamination. To the best of our knowledge, this was the first spectral survey of methyl acetylene toward the source, resulting in interesting insights on the physics of the source. The spectral analysis was performed through rotational diagrams, assuming LTE, from which we derived an averaged excitation temperature of $\sim$50 K for an extended emission. Thus, the bulk of the CH$_3$CCH emission is likely originated from a warm and extended gas, associated with the ambient core region of G331. Moreover, we obtained $N$(CH$_3$CCH)=7.5$\times$10$^{15}$cm$^{−2}$, X[CH$_3$CCH/H$_2$]~$\approx$~(0.77~–~2.8)~$\times$~10$^{−8}$ and X[CH$_3$CCH/CH$_3$OH]~$\approx$~0.42$~\pm$~0.05 from the observations, which are consistent with other single-dish observational works toward massive star-forming regions and suggest that CH$_3$CCH is a zeroth-generation molecule with similar chemistry throughout these environments. 

The $K$=2/$K$=3 line-intensity ratios of transitions within a given K-ladder are strongly negatively correlated with $J_u$, which firmly suggests that the emission is arising from a region with a temperature gradient. An analysis of the rotational diagram separated into two subsections provides upper and lower limits of, respectively, $\sim$60 and $\sim$45 K for the temperature gradient. This approach enables us to assess the small-scale structure of the source with data from single-dish facilities. For a thorough analysis of the temperature profile, however, it is imperative to observe a wide range of CH$_3$CCH lines with a high angular resolution, by means of interferometry instruments such as ALMA.

The line widths of transitions with the same $K$ quantum number are strongly correlated to their rest frequencies and peak velocities. Thus, we infer that the warmer regions, traced by the higher-frequency lines, are also associated with higher turbulence effects. Moreover, the $K$=0 transitions seem to be tracing a slightly different region than the rest of the CH$_3$CCH lines, as indicated by its different kinematic signature. Surveys of this species with higher angular resolution are also fundamental to further explore the origin of this differentiation.

A chemical model of CH$_3$CCH in G331, comprising of a dark-cloud phase followed by a hot-core collapse, predicts CH$_3$CCH abundances that agree with our observations for timescales of $t\gtrsim10^{3}$ yr and densities of $n_{H_2}\sim5\times10^6$ cm$^{-3}$. This is in line with the expected age and size of this source.

\acknowledgments
This publication is based on data acquired with the Atacama Pathfinder Experiment (APEX) under programme IDs C-094.F-9709B-2014, C-097.F-9710A-2016, C-099.F-9702A- 2017 and C-0102.F-9702B-2018. APEX is a collaboration between the Max-Planck-Institut fur Radioastronomie, the European Southern Observatory, and the Onsala Space Observatory. We thank the APEX staff for their helping during the observations and the anonymous referee for their constructive criticism that definitely improved this work. L.B. and R.F. acknowledge support from ANID project Basal AFB-170002. EM acknowledges support from the Brazilian agencies FAPESP, grant 2014/22095-6, and CNPq, grant 150465/2019-0. M.M. acknowledges support from ANID, Programa de Astronomía - Fondo ALMA-CONICYT, project 3119AS0001. RF acknowledge the support of ANID through the ALMA-CONICYT project 31180005. Finally, we would like to acknowledge Dr. Colin Western, who sadly passed away recently, for creating the PGOPHER Software. His contributions to the scientific community will remain evermore.

%

\vspace{5mm}
\facilities{Atacama Pathfinder EXperiment, APEX telescope.}


\software{CASSIS \citep{Vastel2015}, GILDAS
\citep{Pety2005,Gildas2013}, NAUTILUS \citep{Ruaud2016}, PGOPHER \citep{Western2016}.}






\bibliography{mybibfile.bib}{}
\bibliographystyle{aasjournal}



\end{document}